\documentclass[a4paper,fleqn,usenatbib]{mnras}

\usepackage[T1]{fontenc}
\usepackage{ae,aecompl}

\usepackage{graphicx}	
\usepackage{amsmath}	 
\usepackage{amssymb}	 

\graphicspath{ {eps_1/} } 

\title[Constraints for CCSN progenitors with Chandra]{Progenitor constraints for core-collapse supernovae from \textit{Chandra} X-ray observations}

\author[Heikkil{\"a} et al.]{T. Heikkil{\"a}$^{1}$, S. Tsygankov$^{1}$, S. Mattila$^{1, 2, 3}$, J.J. Eldridge$^{4}$, M. Fraser$^{3}$, J. Poutanen$^{1}$\\
$^{1}$Tuorla Observatory, Department of Physics and Astronomy, University of Turku, V\"ais\"al\"antie 20, FI-21500 Piikki\"o, Finland \\
$^{2}$Finnish Centre for Astronomy with ESO (FINCA), University of Turku,  V\"ais\"al\"antie 20, FI-21500 Piikki\"o, Finland \\
$^{3}$Institute of Astronomy, University of Cambridge, Madingley Road, Cambridge CB3 0HA, UK\\
$^{4}$Department of Physics, University of Auckland, Private Bag 92019, Auckland, New Zealand}

\pubyear{2016}

\begin{document}
\label{firstpage}
\pagerange{\pageref{firstpage}--\pageref{lastpage}}
\maketitle

\begin{abstract}
The progenitors of hydrogen-poor core-collapse supernovae (SNe) of types Ib, Ic and IIb are believed to have shed their outer hydrogen envelopes either by extremely strong stellar winds, characteristic of classical Wolf-Rayet stars, or by binary interaction with a close companion star. The exact nature of the progenitors and the relative importance of these processes are still open questions. One relatively unexplored method to constrain the progenitors is to search for high-mass X-ray binaries (HMXB) at SN locations in pre-explosion X-ray observations. In a HMXB, one star has already exploded as a core-collapse SN, producing a neutron star or a stellar-mass black hole. It is likely that the second star in the system will also explode as a supernova, which should cause a detectable long-term change in the system's X-ray luminosity. In particular, a pre-explosion detection of a HMXB coincident with a SN could be informative about the progenitor's nature. In this paper we analyze pre-explosion ACIS observations of 18 nearby type Ib, Ic and IIb supernovae from the \textit{Chandra} X-ray observatory public archive. Two sources that could potentially be associated with the supernova are identified in the sample. Additionally we make similar post-explosion measurements for 46 SNe. Although our modelling indicates that progenitor systems with compact binary companions are probably quite rare, studies of this type can in the future provide more stringent constraints as the number of discovered nearby SNe and suitable pre-explosion X-ray data are both increasing.
\end{abstract}

\begin{keywords}
supernovae: general -- supernovae: individual: SN 2004gt -- stars: Wolf-Rayet -- binaries: close -- X-rays: binaries -- stars: evolution.
\end{keywords}



\section{Introduction} \label{sect_intro}
Supernovae are broadly divided into two classes based on the presence (type II) or absence (type I) of strong hydrogen features in their spectrum \citep[e.g.][]{ref22}. SNe other than type Ia are believed to be the result of core-collapse explosions of massive stars at the end of their lifespan. Among hydrogen-poor core-collapse supernovae (CCSN), the lack of H-features in their spectra is attributed to these stars having lost their hydrogen-envelope prior to the explosion, and hence are collectively referred to as stripped-envelope CCSNe. These are supernova types Ib, Ic and also IIb (which, while they show H-features initially, rapidly lose them and begin to display He-features instead). There are believed to be two primary channels to produce stripped-envelope CCSNe, both of which are likely to be responsible for producing some fraction of these SNe. Previously, the preferred model was that the progenitors were single Wolf-Rayet stars with initial masses above $\sim$25M$_{\odot}$, characterised by high mass-loss rates, resulting in their envelopes being lost through radiatively driven winds \citep[see e.g.][for review]{crowther}. However, recently the alternative mechanism for producing a stripped-envelope SN progenitor, binary interaction, has become the more favored explanation. In an interacting binary scenario, a progenitor that is not sufficiently massive to enter a WR-stage by itself could nonetheless lose its hydrogen envelope either by Roche-lobe overflow to a close companion star or via common-envelope evolution \citep*{ref5}. In fact, based on the studies by e.g. \citet{ref5}, \citet{vanbeveren_1998}, \citet{ref9} and \citet{ref4}, the high relative rate\footnote{According to a study by \citet{ref4} (based on a sample of 100 nearby SNe) 26 percent of CCSNe are of type Ibc and 12 percent type IIb in a volume limited sample (the remaining 62 percent are non-stripped-envelope CCSNe such as type IIP).} of type Ibc SNe could not be reconciled with a single-star scenario where stellar wind dominates the removal of the envelope, and that classical WR-stars could not account for more than half of the observed type Ibc SNe. Instead, the interacting binary scenario was found to be the more likely mechanism for most type Ib and IIb (and large fraction of Ic) SN progenitors. Furthermore, \citet{lyman} analyzed the ejecta-masses of 36 stripped-envelope CCSNe and found that all but three of them were inconsistent with massive progenitor stars, which further supports the likelihood of less massive binary progenitors and suggests this might even be the dominant progenitor mechanism for stripped-envelope CCSNe. It should also be noted that there is a distinct lack of pre-explosion high-mass SN progenitor star detections (with initial mass above $\sim$18M$_{\odot}$) \citep[][]{newsmartt}. It has been suggested that this lack is caused by failed SNe where the core of the star collapses directly into a black hole without detectable SN \citep*[see e.g.][]{gerke_dsn, reynolds_dsn}. This would reduce the population of massive stars available as potential single WR-star SN progenitors. For a more detailed discussion on the various progenitor models, see eg. the recent review by \citet{yoon}.

While a number of SN type II-P progenitor stars have been directly identified in optical pre-explosion observations \citep[see][for review]{ref3}, only three type IIb progenitors have been identified: SN 2008ax \citep{crockett}, SN 1993J \citep*{ref20, maund_smartt} and SN 2011dh \citep{ref21, vandyk_2011dh}, and only one type Ibc progenitor with the recent discovery of the progenitor candidate for the type Ib SN iPTF13bvn by \cite{ref23}. The progenitor of SN 2008ax is the only one of these where a single WR-star progenitor cannot be ruled out, although a binary system seems to be the more likely possibility (see discussion in \cite{ref3} and also \cite{crockett} and \cite{roming}). The progenitor of SN 1993J has been characterised by direct pre-explosion imaging to have been a red supergiant in a binary system. The progenitor of SN 2011dh was likewise identified by direct observation as a yellow supergiant in a probable binary system \citep{maund_smartt, ref21, folatelli}. The type Ib SN iPTF13bvn is also now believed to have had a binary progenitor, the single massive WR-star scenario having been ruled out by \citet{ref24}, \citet{bersten}, \citet{ref25} and \citet{nebular_phase}. Studies of the observed properties of type IIb SNe have also been used to constrain their progenitors. For example, \citet{ref16} found the progenitor of the type IIb SN 2011hs to be consistent with a supergiant star with an initial mass less than 20M$_{\odot}$ as also found for SNe 1993J and 2011dh. Based on radio-observations, \citet{ref17} suggest the progenitor to the type IIb SN 2001ig was a WR-star in a binary, and the likely companion star has been identified as a late-B through late-F type supergiant \citep*{ref19}.

In X-ray frequencies, there has been some effort to identify or constrain the properties of stripped-envelope CCSN progenitors. For example, \citet{ref7} and more recently \citet{ref8} studied the shock-breakout of SN 2008D and found its progenitor to be consistent with a WR-star. In a case study for the type Ib SN 2010O in Arp299, \citet{ref2} used archival pre-explosion \textit{Chandra} data and claimed the detection of a variable X-ray source coincident with the supernova by comparing two observations taken at different times in order to estimate the flux. They suggested this might be indicative of the SN having a high-mass X-ray binary (HMXB) as a progenitor. A HMXB is a binary system consisting of a massive star which is losing mass to a compact companion, with accompanying X-ray emission. Similarly, \citet{ref26} noted the coincidence of the type Ib SN 2009jf with the position of the ultraluminous X-ray source (ULX) CXOU J230453.0+121959, suggesting the possibility that the ULX may be a HMXB progenitor for the SN although they also note that a chance superposition is not unlikely. In this paper, we aim to extend the search for possible HMXB-progenitors for stripped-envelope CCSNe in X-rays. Nielsen et al. recently carried out two studies in which they measured the X-ray luminosity at the sites of 13 type Ia SNe, establishing upper limits for their progenitors \citep*[see][]{ref1, ref37}. Here we make use of the overall methods of these two studies, but using a sample of supernovae of types Ib, Ic and IIb instead of Ia.

The type IIb SNe with pre-explosion progenitor detections (SN 1993J, SN 2008ax and SN 2011dh, as well as the SNe with indirect progenitor constraints, SN 2001ig and SN 2011hs) were also included in this study for comparison. For iPTF13bvn no suitable \textit{Chandra} observations were found and it was therefore excluded. We mainly focus on pre-explosion observations. Based on observed SN X-ray lightcurves the X-ray luminosity of a CCSN can be expected to remain relatively high for many years after the explosion due to the expansion of the SN ejecta into the circumstellar material \citep[see e.g.][]{ref15}, which is the principal source of X-rays during this phase. However, we have also included a large sample of post-explosion results in order to present a more complete picture. Although the main intent is to search for the progenitor of the second supernova in a system that already contains a compact object (neutron star or a stellar-mass black hole), it is quite possible that some of the SNe studied here could also have been the initial SN in the close binary system that resulted in a formation of a HMXB. Some of the SNe in our sample already had previously published luminosities or upper limits derived from the same post-explosion data. In particular, \citet{ref34} reported X-ray luminosities or their upper limits from post-explosion observations of 100 core-collapse SNe of all types from \textit{Chandra}, \textit{XMM-Newton} and \textit{Swift} to study a link between X-ray luminosity and the rotation speeds of pulsars at birth. However, because these previously published values are typically measured using a different selection of observations, energy range, assumed spectral model or aperture, they are not directly comparable to the results presented in this study.

In this paper we use the following structure: In section \ref{sect_obs}, we cover our sample selection and the observations used in this study. In section \ref{sect_data_analysis}, we explain the processes we used for our data analysis, describe the spectral models used, and explain how the results were obtained. In section \ref{sect_results}, we examine the results in detail, consider the theoretical fraction of X-ray bright progenitor systems and present a prediction for the upper limits that can be obtained for future nearby SNe. Finally in section \ref{sect_discussion} we discuss some of the implications these results have for future studies of this type, followed by brief conclusions in section \ref{sect_conclusions}.

\section{Observations} \label{sect_obs}
\textit{Chandra X-ray observatory}, launched in 1999, carries two instruments (HRC and ACIS), and has a spatial resolution of $\sim$0.5 arcsec, although the PSF depends highly on photon energy and distance from the optical axis. Only ACIS observations were included in this study. ACIS is sensitive in the 0.3-8 keV energy range. The precision of the absolute astrometry of \textit{Chandra} data is sufficient for our study, with coordinate offsets being $\sim$1 arcsec or less in most instances\footnote{For details, see http://cxc.harvard.edu/cal/ASPECT/celmon/}.

We selected from the Asiago supernova catalog \citep{asiago} all core-collapse supernovae (as of 2015 May 27) of types Ib, Ic and and IIb discovered after 1980 to ensure reasonably accurate astrometry. We then cross-correlated the known supernova positions with \textit{Chandra} archival data to find observations with ACIS-data from either before or after the supernova explosion. We removed from the sample all SNe with distance more than 100 Mpc. Some SNe were also removed from the sample due to data quality (SN near or outside detector edge, SN more than 10 arcmin from optical axis). To confirm that we were not missing a number of supernovae without an IAU SN designation we also checked the ''Bright Supernova Archive''\footnote{http://www.rochesterastronomy.org/snimages/} for objects between 2013-2015 fulfilling the above criteria. A total of 31 such SNe were found, none of which had suitable ACIS-data. This way, we found that suitable archival ACIS-data was available for approximately 18 percent of SNe within our search criteria. In total, 57 SNe were selected for the sample, of which 11 had only pre-explosion data available, 7 had both pre- and post-explosion data and the remaining 39 post-explosion data only. The full list of all supernovae included in the study and the \textit{Chandra} observations used for each target are presented in Table \ref{obstable}, along with supernova optical maximum or discovery dates (from Asiago) and the combined exposure times for both pre- and post-explosion categories. 

All SN coordinates from Asiago were compared to both the Unified Supernova Catalog (USC) \citep*{lennarz} and the Sternberg Supernova Catalog (SSC) \citep*{sternberg}. Whenever there were significant differences in the SN positions between Asiago and the other two catalogs\footnote{Differences less than $\sim$1 arcsec between the catalogs were not considered significant. Overall, the more recent USC was considered more accurate, and in case of discrepancies, if more accurate astrometry was not available we adopted the USC position instead.}, we attempted to find a different source (see Tables \ref{blocktable_pre} and \ref{blocktable_post}) for the position. In all cases, adjustments were made if more accurate astrometry was available, even if the catalogs were in agreement.

In cases where multiple \textit{Chandra} observations were found for a particular SN, all of the available data were first aligned. The data were then divided into pre- and post-explosion datasets for each SN. If any observation was made within several months of the SN date listed in Asiago catalog\footnote{Asiago catalog max. epoch: The date of optical maximum or discovery.} and there was a possibility of incorrectly assigning an observation to pre- or post-explosion dataset, we ensured the categorization was correct by searching for (or narrowing down) the actual SN explosion date wherever possible. All pre- or post-explosion observations were then combined together to obtain the best possible signal-to-noise (S/N) ratio. Some observations were excluded however, either because their relatively short exposure times of $<$ 5 ks would not have made any significant contribution to the total exposure time of the combined observation, or any data that were otherwise unsuitable, such as observations taken in \textit{Chandra's} continuous-clocking mode. The observations, whether combined or single, were then used to obtain the flux and luminosity at the position of the SN before and/or after the explosion, or an upper limit where no source was detected (as was the case for most SNe).

\begin{table*}
 \centering
 \begin{minipage}{180mm}
   \caption{List of all sample SNe and available \textit{Chandra} observations}\label{obstable}
  \begin{tabular}{@{}lllcc@{}}
  \hline
  Name & \textit{Chandra} observation ID$^\mathrm{A}$ & Epoch$^\mathrm{B}$ & Exposure & Exposure \\
  & & & (pre-SN)(ks) & (post-SN)(ks) \\
  \hline
SN 1983I & 859, 2148, 10403, 10404, 10775, 10776, 10777, 10778, 10779, 10780, 10781, 10782 & 1983 Apr 25 & - & 444.6\\
& 10801, 10824 & \\
SN 1983N & 793, 2064, 12992, 12993, 12994, 12995, 12996, 13202, 13241, 13248, 14332, 14342 & 1983 July 15 & - & 791.3\\
SN 1983V & 3554, 6868, 6869, 6870, 6871, 6872, 6873, 13920, 13921 & 1983 Dec 4 & - & 302.2\\
SN 1984L & 7861 & 1984 Aug 20 & - & 5.1\\
SN 1985F & 7147 & 1984 June 12 & - & 9.4\\
SN 1990U & 10120, 11230 & 1990 Aug 2 & - & 35.4\\
SN 1991N & 2939 & 1991 Apr 2 & - & 48.2 \\
SN 1993J & 735, 5600, 5601, 5935, 5936, 5937, 5938, 5939, 5940, 4941, 5942, 5943, 5944 & 1993 Mar 30 & - & 752.3\\
& 5945, 5946, 5947, 5948, 5949, 6174, 6346, 6347, 6892, 6893, 6894, 6895, 6896 \\
&  6897, 6898, 6899, 6900, 6901, 9122, 12301 & \\
SN 1994I & 354, 1622, 3932, 12562, 12668, 13812, 13813, 13814, 13815, 13816, 15496, 15553 & 1994 Apr 9 & - & 856.1 \\
SN 1996D & 15050 & 1996 Feb 9 & - & 16.0\\
SN 1996N & 16350 & 1996 Mar 12 & - & 48.7\\
SN 1996aq & 11229 & 1996 Aug 17 & - & 7.0 \\
SN 1997X & 4061 & 1997 Feb 1 & - & 10.8\\
SN 1998T & 1641, 6227, 15077, 15619 & 1998 Mar 3 & - & 125.6\\
SN 1998bo & 11505 & 1998 Apr 22 & - & 20.9\\
SN 1998bw & 1956 & 1998 May 10 & - & 48.3\\
SN 1999dn & 4800 & 1999 Aug 27 & - & 60.1\\
SN 1999eh & 9104 & 1999 Oct 12 & - & 18.1\\
SN 1999ex & 10392 & 1999 Nov 15 & - & 12.3\\
SN 2000cr & 10395 & 2000 June 25 & - & 16.1\\
SN 2000ds & 9528 & 2000 Oct 10 & - & 65.5\\
SN 2001ci & 2038 & 2001 Apr 25$^\mathrm{C}$ & 26.9 & -\\
SN 2001ig & 3495, 3496 & 2001 Dec 10 & - & 47.9 \\
SN 2002hf & 2244 & 2002 Oct 29 & 9.2 & -\\
SN 2003L & 4417 & 2003 Jan 12 & - & 30.4\\
SN 2003bg & 3870, 3871, 7605 & 2003 Mar 19 & - & 127.3\\
SN 2003is & 16577 & 2003 Oct 14 & - & 9.8\\
SN 2004C & 4659, 4660, 7607 & 2004 Jan 12 & - & 139.0\\
SN 2004dk & 11226 & 2004 Aug 20 & - & 8.1\\
SN 2004gn & 4017 & 2004 Dec 1 & 5.0 & -\\
SN 2004gt & 315, 3040, 3041, 3042, 3043, 3044, 3718 & 2004 Dec 12 & 425.2 & -\\
SN 2005U & 1641, 6227, 15077, 15619 & 2005 Jan 30 & 24.9 & 100.7\\
SN 2005at & 15384 & 2005 Mar 5 & - & 52.8\\
SN 2005cz & 6785 & 2005 July 17 & - & 15.2\\
SN 2006ep & 7608, 9098 & 2006 Aug 30 & - & 8.0\\
SN 2007Y & 15392, 16487 & 2007 Mar 3 & - & 49.7\\
SN 2007gr & 387, 9579 & 2007 Aug 28 & 2.4 & 19.7\\
SN 2007ke & 908, 11717, 12016, 12017, 12018 & 2007 Sept 24  & 48.5 & 135.5\\
SN 2007kj & 6978, 8491 & 2007 Oct 2 & 46.3 & -\\
SN 2008D & 9104 & 2008 Jan 28$^\mathrm{D}$ & - & 18.1\\
SN 2008ax & 1579, 4725, 4726 & 2008 Mar 22 & 98.9 & -\\
SN 2008bo & 9105, 10118 & 2008 Mar 31 & - & 19.4\\
SN 2009bb & 10140 & 2009 Mar 21 & - & 10.0\\
SN 2009dt & 10265 & 2009 Apr 28 & 5.1 & -\\
SN 2009jf & 10120, 11230 & 2009 Sept 27$^\mathrm{E}$ & 25.2 & 10.2\\ 
SN 2009mk & 11236 & 2009 Dec 15 & - & 10.0\\
SN 2010O & 1641, 6227, 15077, 15619 & 2010 Jan 24 & 35.2 & 90.4\\
SN 2011dh & 354, 1622, 3932, 12562, 12668, 13812, 13813, 13814, 13815, 13816, 15496, 15553 & 2011 June 1 & 90.8 & 765.3 \\
SN 2011ei & 12669 & 2011 July 25 & - & 10.0\\
SN 2011hs & 3947 & 2011 Nov 12 & 55.7 & -\\
SN 2012ap & 13785 & 2012 Feb 18 & - & 9.9\\
SN 2013ak & 14795 & 2013 Mar 9 & - & 9.9\\
SN 2013dk & 315, 3040, 3041, 3042, 3043, 3044, 3718 & 2013 June 22 & 425.2 & -\\
SN 2013ff & 11776 & 2013 Aug 31 & 30.1 & -\\
SN 2013ge & 16556 & 2013 Nov 8 & - & 18.8\\
SN 2014C & 2198, 17569, 17570 & 2014 Jan 5 & 30.1 & 19.8\\
SN 2014L & 7863 & 2014 Jan 26 & 5.1 & -\\
   \hline
 \end{tabular}
 \newline
 [A] Only those \textit{Chandra} OBSIDs are listed which were used for the measurements.
 [B] From Asiago Max. epoch: The date of optical maximum or discovery.
 [C] Explosion epoch narrowed down from \citet{ref40} to ensure the observation is correctly categorized.
 [D] Explosion epoch verified from \citet{ref6} to ensure the observation is correctly categorized.
 [E] Explosion epoch verified from \citet{ref10} to ensure the observations are correctly categorized.
 \end{minipage}
\end{table*}

\begin{table*}
 \centering
  \begin{minipage}{180mm}
  \caption{Models used for the study}\label{models}
  \begin{tabular}{@{}clllcc@{}}
  \hline
  Ref.no & Description & Model & Parameters & $\langle E_{\gamma} \rangle$ (keV)$^\mathrm{A}$ & $C_\mathrm{abs}$ $^\mathrm{B}$\\
  \hline
  1 & Soft-state black-hole (low abs.) & Absorbed blackbody & $T=1$ keV, $N_\mathrm{H}=0.5\times10^{22}$ cm$^{-2}$ & 2.48 & 1.15\\
  2 & Soft-state black-hole (high abs.) & Absorbed blackbody & $T=1$ keV, $N_\mathrm{H}=2.0\times10^{22}$ cm$^{-2}$ & 3.10 & 1.47\\
  3 & Accreting pulsar (low abs.) & Absorbed powerlaw & $\Gamma=0.5$, $N_\mathrm{H}=0.5\times10^{22}$ cm$^{-2}$ & 2.81 & 1.11\\
  4 & Accreting pulsar (high abs.) & Absorbed powerlaw & $\Gamma=0.5$, $N_\mathrm{H}=2.0\times10^{22}$ cm$^{-2}$ & 3.64 & 1.30\\ 
  5 & Hard-state black-hole (low abs.) & Absorbed powerlaw & $\Gamma=1.5$, $N_\mathrm{H}=0.5\times10^{22}$ cm$^{-2}$ & 2.00 & 1.35\\
  6 & Hard-state black-hole (high abs.) & Absorbed powerlaw & $\Gamma=1.5$, $N_\mathrm{H}=2.0\times10^{22}$ cm$^{-2}$ & 3.02 & 1.80\\
   \hline
 \end{tabular}\newline
 [A] Average photon energy for the model.\newline
 [B] Correction coefficient for absorption.
 \end{minipage}
\end{table*}

\section{Data analysis} \label{sect_data_analysis}
We calculated luminosity for three simple spectral models approximating common properties of high-mass X-ray binaries \citep{ref27, ref11}: a blackbody at the temperature of 1 keV (a black hole in soft state; accretion disk), a powerlaw with photon index $\Gamma=$0.5 (accreting pulsar with high magnetic field), and powerlaw with photon index $\Gamma=$1.5 (a black hole in hard state). All three models were further combined with a model of interstellar absorption (XSPEC phabs\footnote{The default values for cross-sections (xsect) and abundances (abund) were used, these being bcmc and angr, respectively.}) with two different absorption values (for a total of six models considered): $0.5\times10^{22}$ cm$^{-2}$ and $2.0\times10^{22}$ cm$^{-2}$. The models and their parameters are also listed in Table \ref{models}. For these models, the main source of absorption is assumed to be the circumstellar material around the progenitor. Assuming a conservative average host galaxy extinction for CCSNe of $A_V\sim$1 \citep{dist11} and converting this into a hydrogen column density according to the formula $N_\mathrm{H}=1.79\times10^{21}A_V$ \citep{predehl} results in a column density of $0.2\times10^{22}$ cm$^{-2}$. This is significantly lower than the absorption values used in our models and we therefore do not consider the effects of host galaxy interstellar gas separately. The one exception to this is SN 2001ci, which has an unusually high $A_V$ of $\sim 5-6$ \citep{ref40} corresponding to $N_\mathrm{H}\sim1.0\times10^{22}$ cm$^{-2}$. We do not account for this higher extinction separately, but note that only the higher-absorption models are realistic for this particular SN. 

The data were processed using \textit{Chandra's} CIAO 4.6 software package. In most cases, only a single observation existed of the given target. For these, prior to analysis, all of the data were reprocessed with the $chandra\_repro$ script to apply standard corrections (using CALDB version 4.6.1.1). In cases where multiple observations of the same area were available, these were first aligned with each other. In most cases, the relative astrometry of the original data was already sufficiently accurate that after reprocessing with $chandra\_repro$, the data could simply be aligned with the $reproject\_obs$ tool. However, in some cases, it was beneficial to apply a correction into the aspect solution of the individual observations to ensure their coordinate grids were better aligned prior to reprojecting their tangent planes. In such cases, a wavelet analysis (with CIAO's $wavdetect$ tool) was first applied to detect a number of sources in each observation. These sources were then matched on the individual observations (using CIAO's $reproject\_aspect$ script) to produce an aspect correction, which was then applied by reprocessing the data with $chandra\_repro$ before finishing the alignment process with $reproject\_obs$.

After alignment and reprocessing, all datasets were divided into pre-explosion and post-explosion data for each supernova. Combined photon-count maps and exposure maps were then made using CIAO's $flux\_obs$ tool for each SN in both categories. The exposure maps are in units of cm$^2$ s and combine instrument quantum efficiency with the exposure time as the telescope is dithered across the sky. They were calculated using spectral weights (produced with $make\_instmap\_weights$) for each of the six models. The energy range used for all models and maps was 0.5-7.0 keV.

For most targets, the flux was measured using a circular aperture with a radius of 2.5 arcsec at the known position of the SN (which contains more than 90 percent of \textit{Chandra's} PSF for sources less than 2 arcmin from the optical axis, see http://cxc.harvard.edu/proposer/POG/html/chap4.html, Fig.4.13). In cases where the optical axis of one or more of the original individual observations was further than 2 arcmin from the SN position, the source aperture radius was increased accordingly (in increments of 0.5 arcsec) so that it contained at least 90 percent of the PSF. Background was measured using an annulus around the source aperture circle, the size of which varied according to target. Generally the outer radius of the annulus was increased either to maintain roughly 2/5 ratio between the source aperture radius and the background annulus outer radius, or increased even further to ensure that at least 10 photons were located within the background area. For a list of the used source aperture and background annulus sizes, see Tables \ref{blocktable_pre} and \ref{blocktable_post}. Any unrelated sources were excluded from both the source apertures and the background annuli. For the purposes of this study, we conservatively consider a source unrelated if it is located further than $\sim$1.5 arcsec from the SN position or is not a point-source. For most pre-explosion targets there were no point-sources closer than 5 arcsec to the SN position. In all but two of the cases where more nearby point-sources existed, the position of the SN was known from literature with sufficient accuracy to rule out any association between such a source and the SN. The two exceptions, SN 2009jf and SN 2004gt, will be discussed in more detail in section \ref{sect_detected}.

Background corrected flux was measured for each source aperture. If no source was found (within $3\sigma$), then a $3\sigma$ upper limit was calculated instead using the number of detected counts inside the source aperture (equation \ref{flux_ulim} below). However, in many cases a short exposure time would result in the number of photons inside the aperture itself to be very low (or even zero), so in cases where there were no visible sources and the number of counts inside the source aperture was $<20$, we used Poissonian statistics to find the matching upper limit at a 99.87 percent confidence level \citep[corresponding to Gaussian $3\sigma$, see e.g.][]{gehrels}, and calculated the $3\sigma$ upper limit using equation \ref{flux_poisson} below instead for these cases. To obtain unabsorbed flux, we calculated an absorption correction coefficient for each model by comparing the absorbed models with the corresponding unabsorbed model in XSPEC. The coefficients are listed in table \ref{models}. Following \citet{ref1}, for any $3\sigma$ detection we calculated the flux from

\begin{equation}\label{flux}
F= \frac{(n-b) \langle E_{\gamma} \rangle}{\zeta}
\end{equation}
where $n$ is the number of photons from the source aperture, $\langle E_{\gamma} \rangle$ is average photon energy for the given model (see table \ref{models}), $\zeta$ is the average value of the exposure map within the source aperture and $b$ is scaled background, the number of background counts $n_{\mathrm{bg}}$ scaled to the size of the source aperture: 
\begin{equation}\label{bgscaled}
b= \frac{n_{\mathrm{bg}} \times A_{\mathrm{src}}}{A_{\mathrm{bg}}},
\end{equation}
where $A_{\mathrm{src}}$ and $A_\mathrm{bg}$ are the surface areas of the source and background regions respectively. The flux upper limit was calculated from
\begin{equation}\label{flux_ulim}
F_\mathrm{UL}= \frac{3\sqrt{n}\langle E_{\gamma} \rangle}{\zeta}
\end{equation}
for high photon count cases. For the low photon count cases, the flux upper limit was instead calculated from
\begin{equation}\label{flux_poisson}
F_\mathrm{UL}=\frac{(\mu -b)\langle E_{\gamma} \rangle}{\zeta},
\end{equation}
where $\mu$ is the 0.9987 confidence-level Poissonian upper limit corresponding to the number of photons within the source aperture, according to \citet{gehrels}. 
Finally, the unabsorbed luminosity for each $F$ and $F_\mathrm{UL}$ was calculated from
\begin{equation}\label{lum}
L= 4 \pi C_\mathrm{abs} F d^{2}
\end{equation}

where $C_\mathrm{abs}$ is the correction coefficient for absorption and $d$ is the distance to the host galaxy. The distances used for calculating the luminosities were obtained via NED\footnote{http://ned.ipac.caltech.edu/} (see Tables \ref{blocktable_pre} and \ref{blocktable_post} for details), or calculated from the Hyperleda\footnote{\citet{refleda}: See http://atlas.obs-hp.fr/hyperleda/} radial velocity corrected for infall on to the Virgo cluster (vvir) using $H_0=70$ km s$^{-1}$Mpc$^{-1}$. 

\section{Results} \label{sect_results}
Basic information about each SN (host galaxy, SN type, position, distance), as well as non-model-specific results of our measurements are presented in Table \ref{blocktable_pre} for those SNe with pre-explosion data, and in Table \ref{blocktable_post} for those with post-explosion data. Targets with both types of data are present in both. The model-specific results -- average exposure map values and luminosity upper limits -- are presented in Table \ref{resultstable_pre1} and Tables \ref{resultstable_post1} and \ref{resultstable_post2} for pre- and post-explosion data respectively. Any measured luminosities for sources that have a flux above the 3$\sigma$ detection threshold and their errors are also included in these tables. The models are numbered according to Table \ref{models}.

\begin{table*}
 \centering
 \begin{minipage}{180mm}
   \caption{SNe with pre-explosion \textit{Chandra} X-ray data and measured photon counts}\label{blocktable_pre}
  \begin{tabular}{@{}llllllll@{}}
  \hline
  Name & Host & SN & Position$^\mathrm{A}$ & Distance$^\mathrm{B}$ & Aperture$^\mathrm{C}$ & Counts & Scaled \\
  & Galaxy & Type & & (Mpc) & & (Aperture) & background\\
  \hline  
SN 2001ci  & NGC 3079 & Ic &  10:01:57.21 +55:41:14.0$^\mathrm{K}$  &  18.1$^\mathrm{D}$  & 2.5"/6" &  4  &  4.1  \\
SN 2002hf  & MCG-05-03-020 & Ic &  00:57:47.74 $-$27:30:21.5$^\mathrm{U}$  &  77.5  & 3.5"/9" &  10  &  9.3  \\
SN 2004gn  & NGC 4527 & Ic &  12:34:12.10 +02:39:34.4$^\mathrm{U}$  &  14.2$^\mathrm{E}$  & 2.5"/25" &  0  &  0.2  \\
SN 2004gt  & NGC 4038 & Ic &  12:01:50.42 $-$18:52:13.5$^\mathrm{M}$  &  20.4$^\mathrm{F}$  & 2.5"/12" &  694  &  206.4  \\
SN 2005U  & Arp 299 & IIb &  11:28:33.13 +58:33:41.3$^\mathrm{L}$  &  48.2  & 2.5"/6" &  15  &  11.1  \\
SN 2007gr  & NGC 1058 & Ic &  02:43:27.98 +37:20:44.7$^\mathrm{U}$  &  9.3$^\mathrm{G}$  & 2.5"/25" &  0  &  0.1  \\
SN 2007ke  & NGC 1129 & Ib &  02:54:23.90 +41:34:16.3$^\mathrm{U}$  &  77.2  & 6.5"/16" &  430  &  467.5  \\
SN 2007kj  & NGC 7803 & Ib &  00:01:19.58 +13:06:30.6$^\mathrm{U}$  &  77.0  & 2.5"/8" &  3  &  1.5  \\
SN 2008ax  & NGC 4490 & IIb &  12:30:40.80 +41:38:14.5$^\mathrm{U}$  &  9.6$^\mathrm{H}$  & 2.5"/7" &  8  &  8.2  \\
SN 2009dt  & IC 5169 & Ic &  22:10:09.27 $-$36:05:42.6$^\mathrm{U}$  &  42.0  & 2.5"/20" &  0  &  0.2  \\
SN 2009jf  & NGC 7479 & Ib &  23:04:52.98 +12:19:59.5$^\mathrm{U}$  &  34.9  & 3"/8" &  105  &  1.7  \\
SN 2010O  & Arp 299 & Ib &  11:28:33.86 +58:33:51.6$^\mathrm{U}$  &  48.2  & 2.5"/6" &  239  &  67.3  \\
SN 2011dh  & NGC 5194 & IIb &  13:30:05.11 +47:10:10.9  &  7.8$^\mathrm{I}$  & 3"/8" &  3  &  8.7  \\
SN 2011hs  & IC 5267 & IIb &  22:57:11.77 $-$43:23:04.8  &  21.8  & 2.5"/10" &  1  &  1.4  \\
SN 2013dk  & NGC 4038 & Ic &  12:01:52.72 $-$18:52:18.3  &  20.4$^\mathrm{F}$  & 2.5"/6" &  202  &  172.7  \\
SN 2013ff  & NGC 2748 & Ic &  09:13:38.88 +76:28:10.8  &  24.8  & 2.5"/8" &  1  &  2.6  \\
SN 2014C  & NGC 7331 & Ib &  22:37:05.60 +34:24:31.9  &  15.1$^\mathrm{E}$  & 2.5"/6" &  5  &  5.8  \\
SN 2014L  & NGC 4254 & Ic &  12:18:48.68 +14:24:43.5  &  17.3$^\mathrm{J}$  & 2.5"/10" &  1  &  1.1  \\
\hline
 \end{tabular}
 \newline
 [A] From Asiago if not indicated otherwise.
 [B] Distances that do not have a reference have been calculated from Hyperleda radial velocity corrected for the infall to the Virgo cluster (vvir) with $H_0 = 70$ km s$^{-1}$Mpc$^{-1}$.
 [C] Radius of source aperture/outer radius of background annulus.
 [D] Average of values from \citet{dist7, dist16} 
 [E] \citet{dist4} 
 [F] \citet*{dist14} 
 [G] \citet{dist3} 
 [H] \citet{dist6} 
 [I] \citet{dist2} 
 [J] Average of values from \citet{dist15, dist16, dist10} 
 [K] Position confirmed from \citet*{vandykpos}.
 [L] Position from \citet*{2005U_pos_Li} (Li position)
 [M] Position derived in this paper, see section \ref{sect_detected}.
 [U] Position confirmed from Unified Supernova Catalog \citep{lennarz}
  
 \end{minipage}
\end{table*}

\begin{table*} 
 \centering
 \begin{minipage}{180mm}
   \caption{SNe with post-explosion \textit{Chandra} X-ray data and measured photon counts}\label{blocktable_post}
  \begin{tabular}{@{}llllllll@{}}
  \hline
  Name & Host & SN & Position$^\mathrm{A}$ & Distance$^\mathrm{B}$ & Aperture$^\mathrm{C}$ & Counts & Scaled \\
  & Galaxy & Type & & (Mpc) & & (Aperture) & background \\
  \hline  
SN 1983I  & NGC 4051 & Ic &  12:03:11.77 +44:31:00.6  &  13.7$^\mathrm{D}$  & 2.5"/8" &  17  &  17.5  \\
SN 1983N  & NGC 5236 & Ib &  13:36:51.24 $-$29:54:02.7$^\mathrm{U}$  &  4.6$^\mathrm{E}$  & 5"/13" &  301  &  264.7  \\
SN 1983V  & NGC 1365 & Ic &  03:33:31.63 $-$36:08:55.0$^\mathrm{U}$  &  19.6$^\mathrm{E}$  & 2.5"/6" &  16  &  12.3  \\
SN 1984L  & NGC 991 & Ib &  02:35:30.52 $-$07:09:30.5  &  20.4  & 2.5"/20" &  0  &  0.2  \\
SN 1985F  & NGC 4618 & Ib &  12:41:33.01 +41:09:05.9$^\mathrm{U}$  &  7.9$^\mathrm{F}$  & 2.5"/15" &  1  &  0.4  \\
SN 1990U  & NGC 7479 & Ib &  23:04:54.92 +12:18:20.1$^\mathrm{U}$  &  34.9  & 2.5"/8" &  3  &  3.2  \\
SN 1991N  & NGC 3310 & Ic &  10:38:46.37 +53:30:04.7$^\mathrm{U}$  &  17.5$^\mathrm{G}$  & 2.5"/6" &  162  &  117.5  \\
SN 1993J  & NGC 3031 & IIb &  09:55:24.77 +69:01:13.7$^\mathrm{O}$  &  3.6$^\mathrm{H}$  & 3"/8" &  9498  &  60.2  \\
SN 1994I  & NGC 5194 & Ic &  13:29:54.12 +47:11:30.4$^\mathrm{P}$  &  7.8$^\mathrm{I}$  & 3"/8" &  362  &  200.9  \\
SN 1996D  & NGC 1614 & Ic &  04:34:00.30 $-$08:34:44.0  &  66.4  & 2.5"/6" &  27  &  15.3  \\
SN 1996N  & NGC 1398 & Ib &  03:38:55.31 $-$26:20:04.1$^\mathrm{U}$  &  17.2  & 2.5"/10" &  0  &  1.0  \\
SN 1996aq  & NGC 5584 & Ic &  14:22:22.72 $-$00:23:23.8$^\mathrm{U}$  &  22.0$^\mathrm{J}$  & 2.5"/20" &  1  &  0.3  \\
SN 1997X  & NGC 4691 & Ib &  12:48:14.28 $-$03:19:58.5  &  16.4  & 2.5"/6" &  1  &  7.1  \\
SN 1998T  & Arp 299 & Ib &  11:28:33.16 +58:33:43.7$^\mathrm{U}$  &  48.2  & 2.5"/6" &  603  &  216.1  \\
SN 1998bo  & ESO185-031 & Ic &  19:57:22.55 $-$55:08:18.4$^\mathrm{U}$  &  67.4  & 10"/30" &  12  &  9.8  \\
SN 1998bw  & E184-G82 & Ic &  19:35:03.31 $-$52:50:44.6$^\mathrm{Q}$  &  33.8  & 2.5"/6" &  83  &  4.3  \\
SN 1999dn  & NGC 7714 & Ib &  23:36:14.81 +02:09:08.4$^\mathrm{R}$  &  40.0  & 2.5"/6" &  8  &  7.7  \\
SN 1999eh  & NGC 2770 & Ib &  09:09:32.67 +33:07:16.9$^\mathrm{U}$  &  29.5  & 2.5"/10" &  1  &  0.7  \\
SN 1999ex  & IC 5179  & Ib &  22:16:07.27 $-$36:50:53.7$^\mathrm{U}$  &  46.3$^\mathrm{K}$  & 2.5"/12" &  3  &  1.0  \\
SN 2000cr  & NGC 5395 & Ic &  13:58:38.37 +37:26:12.9$^\mathrm{U}$  &  52.7  & 2.5"/8" &  0  &  1.7  \\
SN 2000ds  & NGC 2768 & Ib &  09:11:36.28 +60:01:43.3$^\mathrm{V}$  &  23.2  & 2.5"/6" &  8  &  4.1  \\
SN 2001ig  & NGC 7424 & IIb &  22:57:30.69 $-$41:02:25.9$^\mathrm{U}$  &  10.8$^\mathrm{L}$  & 2.5"/7" &  38  &  1.1  \\
SN 2003L  & NGC 3506 & Ic &  11:03:12.33 +11:04:38.3$^\mathrm{U}$  &  92.2  & 2.5"/6" &  43  &  6.2  \\
SN 2003bg  & M-05-10-15 & Ic &  04:10:59.42 $-$31:24:50.3$^\mathrm{U}$  &  16.3  & 2.5"/7" &  632  &  7.1  \\
SN 2003is  & MCG+07-40-03  & Ic & 19:21:08.00 +43:19:35.4$^\mathrm{U}$  &  81.9  & 5"/22" &  0  &  0.9  \\
SN 2004C  & NGC 3683 & Ic &  11:27:29.72 +56:52:48.2$^\mathrm{U}$  &  27.8  & 2.5"/6" &  136  &  30.3  \\
SN 2004dk  & NGC 6118 & Ib &  16:21:48.93 $-$02:16:17.3$^\mathrm{U}$  &  23.5  & 2.5"/12" &  7  &  0.6  \\
SN 2005U  & Arp 299 & IIb &  11:28:33.13 +58:33:41.3$^\mathrm{S}$  &  48.2  & 2.5"/6" &  90  &  53.4  \\
SN 2005at  & NGC 6744 & Ic &  19:09:53.57 $-$63:49:22.8$^\mathrm{U}$  &  9.0  & 2.5"/10" &  1  &  0.8  \\
SN 2005cz  & NGC 4589 & Ib &  12:37:27.85 +74:11:24.5$^\mathrm{U}$  &  32.6  & 2.5"/8" &  0  &  1.6  \\
SN 2006ep  & NGC 0214 & Ib &  00:41:24.88 +25:29:46.7$^\mathrm{U}$  &  66.1  & 2.5"/15" &  0  &  0.3  \\
SN 2007Y  & NGC 1187 & Ib &  03:02:35.92 $-$22:53:50.1$^\mathrm{U}$  &  17.3  & 3.5"/10" &  4  &  1.7  \\
SN 2007gr  & NGC 1058 & Ic &  02:43:27.98 +37:20:44.7$^\mathrm{U}$  &  9.3$^\mathrm{M}$  & 2.5"/25" &  3  &  0.5  \\
SN 2007ke  & NGC 1129 & Ib &  02:54:23.90 +41:34:16.3$^\mathrm{U}$  &  77.2  & 6.5"/16" &  1182  &  1215.6  \\
SN 2008D  & NGC 2770 & Ib &  09:09:30.65 +33:08:20.3$^\mathrm{U}$  &  29.5  & 2.5"/10" &  11  &  0.7  \\
SN 2008bo  & NGC 6643 & IIb &  18:19:54.41 +74:34:21.0  &  25.4  & 2.5"/8" &  1  &  0.8  \\
SN 2009bb  & NGC 3278 & Ic &  10:31:33.88 $-$39:57:30.0$^\mathrm{T}$  &  40.0  & 2.5"/8" &  10  &  1.7  \\
SN 2009jf  & NGC 7479 & Ib &  23:04:52.98 +12:19:59.5$^\mathrm{U}$  &  34.9  & 3"/8" &  76  &  1.5  \\
SN 2009mk  & E293-G34 & IIb &  00:06:21.37 $-$41:28:59.8$^\mathrm{U}$  &  18.7  & 2.5"/15" &  2  &  0.5  \\
SN 2010O  & Arp 299 & Ib &  11:28:33.86 +58:33:51.6$^\mathrm{U}$  &  48.2  & 2.5"/6" &  759  &  244.3  \\
SN 2011dh  & NGC 5194 & IIb &  13:30:05.11 +47:10:10.9  &  7.8$^\mathrm{I}$  & 3"/8" &  1193  &  49.1  \\
SN 2011ei  & NGC 6925 & IIb &  20:34:22.62 $-$31:58:23.6  &  38.3  & 2.5"/8" &  1  &  1.3  \\
SN 2012ap  & NGC 1729 & Ic &  05:00:13.72 $-$03:20:51.2  &  50.5  & 2.5"/20" &  2  &  0.3  \\
SN 2013ak  & E430-G20 & IIb &  08:07:06.69 $-$28:03:10.1  &  11.6$^\mathrm{N}$  & 2.5"/15" &  30  &  0.4  \\
SN 2013ge  & NGC 3287 & Ic &  10:34:48.46 +21:39:41.9  &  19.7  & 2.5"/12" &  0  &  0.5  \\
SN 2014C  & NGC 7331 & Ib &  22:37:05.60 +34:24:31.9  &  15.1  & 2.5"/6" &  515  &  6.0  \\
  \hline

 \end{tabular}
\newline

 [A] From Asiago if not indicated otherwise.
 [B] Distances that do not have a reference have been calculated from Hyperleda radial velocity corrected for the infall to the Virgo cluster (vvir) with $H_0 = 70$ km s$^{-1}$Mpc$^{-1}$.
 [C] Radius of source aperture/outer radius of background annulus.
 [D] Average of values from \citet{dist7, dist8, dist9, dist10}. 
 [E] \citet{dist4} 
 [F] \citet*{dist5} 
 [G] \citet*{dist12} 
 [H] \citet{dist1} 
 [I] \citet{dist2} 
 [J] \citet{dist14} 
 [K] Average of values from \citet{dist17, dist18, dist19, dist20, dist21, dist22, dist23, dist24, dist25, dist26, dist27}. 
 [L] \citet{dist11} 
 [M] \citet{dist3} 
 [N] Average value from \citet{dist13}. 
 [O] Position from \citet{pos_1993J}.
 [P] Position from \citet{pos_1994I}.
 [Q] Position from \citet{pos_1998bw}.
 [R] Position confirmed from \citet{vandykpos}.
 [S] Position from \citet{2005U_pos_Li}
 [T] Position confirmed from \citet{pos_2009bb}.
 [U] Position confirmed from Unified Supernova Catalog \citep{lennarz}.
 [V] Position obtained from Unified Supernova Catalog \citep{lennarz}.

 \end{minipage}
\end{table*}

\begin{table*} 
 \centering
 \begin{minipage}{180mm}
 \caption{X-ray luminosities and upper limits of X-ray luminosity (pre-explosion)}\label{resultstable_pre1}
  \begin{tabular}{@{}lllllllllllll@{}}

  \hline \hline
 & \multicolumn{2}{@{}c}{ (1) Soft-state BH (low abs.)}  & \multicolumn{2}{@{}c}{ (2) Soft-state BH (high abs.)} & \multicolumn{2}{@{}c}{ (3) Accr. pulsar (low abs.)} \\

Name & Exp. map av. & Luminosity & Exp. map av. & Luminosity & Exp. map av. & Luminosity \\
&  value (cm$^2$s) & (erg s$^{-1}$) &  value (cm$^2$s) & (erg s$^{-1}$) &  value (cm$^2$s) & (erg s$^{-1}$) \\
\hline
2001ci$^\mathrm{A}$ & 1.17$\times10^7$ & $<$ 1.6$\times10^{38}$ & 1.06$\times10^7$ & $<$ 2.8$\times10^{38}$ & 1.04$\times10^7$ & $<$ 1.9$\times10^{38 }$ \\
2002hf & 3.83$\times10^6$ & $<$ 1.2$\times10^{40}$ & 3.48$\times10^6$ & $<$ 2.2$\times10^{40}$ & 3.41$\times10^6$ & $<$ 1.5$\times10^{40}$  \\
2004gn & 2.12$\times10^6$ & $<$ 3.3$\times10^{38}$ & 1.93$\times10^6$ & $<$ 5.8$\times10^{38}$ & 1.89$\times10^6$ & $<$ 4.1$\times10^{38}$  \\
2004gt$^\mathrm{B}$ & 1.80$\times10^8$ & $($6.1$\pm$0.3$)\times10^{38}$ & 1.64$\times10^8$ & $($10.8$\pm$0.6$)\times10^{38}$ & 1.61$\times10^8$ & $($7.6$\pm$0.4$)\times10^{38}$  \\
2005U & 7.25$\times10^6$ & $<$ 3.4$\times10^{39}$ & 6.86$\times10^6$ & $<$ 5.8$\times10^{39}$ & 6.72$\times10^6$ & $<$ 4.1$\times10^{39}$  \\
2007gr & 9.29$\times10^5$ & $<$ 3.3$\times10^{38}$ & 8.32$\times10^5$ & $<$ 5.9$\times10^{38}$ & 8.24$\times10^5$ & $<$ 4.1$\times10^{38}$  \\
2007ke & 1.57$\times10^7$ & $<$ 1.3$\times10^{40}$ & 1.48$\times10^7$ & $<$ 2.2$\times10^{40}$ & 1.44$\times10^7$ & $<$ 1.5$\times10^{40}$  \\
2007kj & 1.37$\times10^7$ & $<$ 2.6$\times10^{39}$ & 1.31$\times10^7$ & $<$ 4.4$\times10^{39}$ & 1.27$\times10^7$ & $<$ 3.1$\times10^{39}$  \\
2008ax & 4.15$\times10^7$ & $<$ 1.5$\times10^{37}$ & 3.78$\times10^7$ & $<$ 2.7$\times10^{37}$ & 3.69$\times10^7$ & $<$ 1.9$\times10^{37}$  \\
2009dt & 1.76$\times10^6$ & $<$ 3.5$\times10^{39}$ & 1.69$\times10^6$ & $<$ 5.9$\times10^{39}$ & 1.63$\times10^6$ & $<$ 4.2$\times10^{39}$  \\
2009jf & 1.03$\times10^7$ & $($6.7$\pm$0.7$)\times10^{39}$ & 9.40$\times10^6$ & $($1.2$\pm$0.1$)\times10^{40}$ & 9.12$\times10^6$ & $($8.3$\pm$0.8$)\times10^{39}$ \\
2010O$^\mathrm{C}$ & 1.09$\times10^7$ & $($2.0$\pm$0.2$)\times10^{40}$ & 1.01$\times10^7$ & $($3.4$\pm$0.3$)\times10^{40}$ & 9.90$\times10^6$ & $($2.4$\pm$0.2$)\times10^{40}$ \\
2011dh & 3.78$\times10^7$ & $<$ 3.5$\times10^{36}$ & 3.44$\times10^7$ & $<$ 6.1$\times10^{36}$ & 3.37$\times10^7$ & $<$ 4.3$\times10^{36}$  \\
2011hs & 2.27$\times10^7$ & $<$ 8.6$\times10^{37}$ & 2.07$\times10^7$ & $<$ 1.5$\times10^{38}$ & 2.02$\times10^7$ & $<$ 1.1$\times10^{38}$  \\
2013dk & 1.79$\times10^8$ & $<$ 5.4$\times10^{37}$ & 1.62$\times10^8$ & $<$ 9.6$\times10^{37}$ & 1.59$\times10^8$ & $<$ 6.7$\times10^{37}$  \\
2013ff & 1.24$\times10^7$ & $<$ 1.7$\times10^{38}$ & 1.14$\times10^7$ & $<$ 3.0$\times10^{38}$ & 1.11$\times10^7$ & $<$ 2.1$\times10^{38}$  \\
2014C & 1.30$\times10^7$ & $<$ 9.8$\times10^{37}$ & 1.18$\times10^7$ & $<$ 1.7$\times10^{38}$ & 1.16$\times10^7$ & $<$ 1.2$\times10^{38}$  \\
2014L & 2.06$\times10^6$ & $<$ 6.2$\times10^{38}$ & 1.89$\times10^6$ & $<$ 1.1$\times10^{39}$ & 1.84$\times10^6$ & $<$ 7.6$\times10^{38}$  \\      
      \hline \hline

 & \multicolumn{2}{@{}c}{(4) Accr. pulsar (high abs.)}  & \multicolumn{2}{@{}c}{(5) Hard-state BH (low abs.)} & \multicolumn{2}{@{}c}{(6) Hard-state BH (high abs.)} \\

Name & Exp. map av. & Luminosity & Exp. map av. & Luminosity & Exp. map av. & Luminosity\\
&  value (cm$^2$s) & (erg s$^{-1}$) &  value (cm$^2$s) & (erg s$^{-1}$) &  value (cm$^2$s) & (erg s$^{-1}$) \\
\hline
2001ci$^\mathrm{A}$ & 9.23$\times10^6$ & $<$ 3.3$\times10^{38}$ & 1.23$\times10^7$ & $<$ 1.4$\times10^{38}$ & 1.05$\times10^7$ & $<$ 3.4$\times10^{38}$  \\
2002hf & 3.04$\times10^6$ & $<$ 2.6$\times10^{40}$ & 3.99$\times10^6$ & $<$ 1.1$\times10^{40}$ & 3.44$\times10^6$ & $<$ 2.6$\times10^{40}$  \\
2004gn & 1.69$\times10^6$ & $<$ 6.9$\times10^{38}$ & 2.20$\times10^6$ & $<$ 3.0$\times10^{38}$ & 1.91$\times10^6$ & $<$ 7.0$\times10^{38}$  \\
2004gt$^\mathrm{B}$ & 1.43$\times10^8$ & $($12.9$\pm$0.7$)\times10^{38}$ & 1.89$\times10^8$ & $($5.6$\pm$0.3$)\times10^{38}$ & 1.62$\times10^8$ & $($13.1$\pm$0.7$)\times10^{38}$ \\
2005U & 6.30$\times10^6$ & $<$ 6.6$\times10^{39}$ & 7.32$\times10^6$ & $<$ 3.2$\times10^{39}$ & 6.78$\times10^6$ & $<$ 7.0$\times10^{39}$  \\
2007gr & 7.24$\times10^5$ & $<$ 7.0$\times10^{38}$ & 9.86$\times10^5$ & $<$ 2.9$\times10^{38}$ & 8.25$\times10^5$ & $<$ 7.1$\times10^{38}$  \\
2007ke & 1.34$\times10^7$ & $<$ 2.5$\times10^{40}$ & 1.59$\times10^7$ & $<$ 1.2$\times10^{40}$ & 1.46$\times10^7$ & $<$ 2.7$\times10^{40}$  \\
2007kj & 1.21$\times10^7$ & $<$ 5.0$\times10^{39}$ & 1.36$\times10^7$ & $<$ 2.5$\times10^{39}$ & 1.29$\times10^7$ & $<$ 5.3$\times10^{39}$  \\
2008ax & 3.30$\times10^7$ & $<$ 3.2$\times10^{37}$ & 4.30$\times10^7$ & $<$ 1.4$\times10^{37}$ & 3.73$\times10^7$ & $<$ 3.2$\times10^{37}$  \\
2009dt & 1.55$\times10^6$ & $<$ 6.6$\times10^{39}$ & 1.74$\times10^6$ & $<$ 3.4$\times10^{39}$ & 1.66$\times10^6$ & $<$ 7.1$\times10^{39}$  \\
2009jf & 8.20$\times10^6$ & $($1.4$\pm$0.1$)\times10^{40}$ & 1.05$\times10^7$ & $($6.2$\pm$0.6$)\times10^{39}$ & 9.27$\times10^6$ & $($1.4$\pm$0.1$)\times10^{40}$ \\
2010O$^\mathrm{C}$ & 9.13$\times10^6$ & $($4.0$\pm$0.4$)\times10^{40}$ & 1.11$\times10^7$ & $($1.9$\pm$0.2$)\times10^{40}$ & 1.00$\times10^7$ & $($4.2$\pm$0.4$)\times10^{40}$ \\
2011dh & 3.01$\times10^7$ & $<$ 7.2$\times10^{36}$ & 3.93$\times10^7$ & $<$ 3.2$\times10^{36}$ & 3.40$\times10^7$ & $<$ 7.4$\times10^{36}$  \\
2011hs & 1.81$\times10^7$ & $<$ 1.8$\times10^{38}$ & 2.35$\times10^7$ & $<$ 7.9$\times10^{37}$ & 2.04$\times10^7$ & $<$ 1.8$\times10^{38}$  \\
2013dk & 1.42$\times10^8$ & $<$ 1.1$\times10^{38}$ & 1.87$\times10^8$ & $<$ 4.9$\times10^{37}$ & 1.60$\times10^8$ & $<$ 1.2$\times10^{38}$  \\
2013ff & 1.00$\times10^7$ & $<$ 3.5$\times10^{38}$ & 1.27$\times10^7$ & $<$ 1.6$\times10^{38}$ & 1.13$\times10^7$ & $<$ 3.6$\times10^{38}$  \\
2014C & 1.03$\times10^7$ & $<$ 2.1$\times10^{38}$ & 1.37$\times10^7$ & $<$ 8.9$\times10^{37}$ & 1.17$\times10^7$ & $<$ 2.1$\times10^{38}$  \\
2014L & 1.66$\times10^6$ & $<$ 1.3$\times10^{39}$ & 2.12$\times10^6$ & $<$ 5.7$\times10^{38}$ & 1.87$\times10^6$ & $<$ 1.3$\times10^{39}$  \\
      \hline \hline
      \end{tabular}
      \newline
      [A] SN 2001ci has an unusually high host galaxy absorption which has not been corrected for here. Therefore, only the higher-absorption models (2, 4 and 6) should be considered realistic.
      [B] The luminosities presented here for SN 2004gt are the source aperture luminosities for the original 0.5-7.0 keV energy range and do not reflect the properties of the source discussed in section \ref{sect_detected}.
      [C] The luminosities presented here for SN 2010O are a combination of multiple different sources found inside the source aperture and do not represent any particular single source, see section \ref{sect_detected}.
            \end{minipage}
      \end{table*}

\begin{table*} 
 \centering
 \begin{minipage}{180mm}
 \caption{X-ray luminosities and upper limits of X-ray luminosity (post-explosion, part 1)}\label{resultstable_post1}
  \begin{tabular}{@{}lllllllllllll@{}}

  \hline \hline

 & \multicolumn{2}{@{}c}{ (1) Soft-state BH (low abs.)}  & \multicolumn{2}{@{}c}{ (2) Soft-state BH (high abs.)} & \multicolumn{2}{@{}c}{ (3) Accr. pulsar (low abs.)} \\

Name & Exp. map av. & Luminosity & Exp. map av. & Luminosity & Exp. map av. & Luminosity \\
&  value (cm$^2$s) & (erg s$^{-1}$) &  value (cm$^2$s) & (erg s$^{-1}$) &  value (cm$^2$s) & (erg s$^{-1}$) \\
\hline
1983I & 5.69$\times10^7$ & $<$ 2.9$\times10^{37}$ & 5.91$\times10^7$ & $<$ 4.5$\times10^{37}$ & 5.67$\times10^7$ & $<$ 3.2$\times10^{37}$  \\
1983N & 3.08$\times10^8$ & $<$ 2.0$\times10^{36}$ & 2.83$\times10^8$ & $<$ 3.4$\times10^{36}$ & 2.73$\times10^8$ & $<$ 2.4$\times10^{36}$  \\
1983V & 6.22$\times10^7$ & $<$ 6.7$\times10^{37}$ & 6.08$\times10^7$ & $<$ 1.1$\times10^{38}$ & 5.86$\times10^7$ & $<$ 7.7$\times10^{37}$  \\
1984L & 2.15$\times10^6$ & $<$ 6.8$\times10^{38}$ & 1.97$\times10^6$ & $<$ 1.2$\times10^{39}$ & 1.91$\times10^6$ & $<$ 8.4$\times10^{38}$  \\
1985F & 3.97$\times10^6$ & $<$ 7.3$\times10^{37}$ & 3.64$\times10^6$ & $<$ 1.3$\times10^{38}$ & 3.54$\times10^6$ & $<$ 9.0$\times10^{37}$  \\
1990U & 1.43$\times10^7$ & $<$ 4.4$\times10^{38}$ & 1.31$\times10^7$ & $<$ 7.7$\times10^{38}$ & 1.27$\times10^7$ & $<$ 5.4$\times10^{38}$  \\
1991N & 2.04$\times10^7$ & $($3.7$\pm$1.0$)\times10^{38}$ & 1.85$\times10^7$ & $($6.4$\pm$1.8$)\times10^{38}$ & 1.81$\times10^7$ & $($4.5$\pm$1.3$)\times10^{38}$ \\
1993J & 1.50$\times10^8$ & $($45.3$\pm$0.5$)\times10^{37}$ & 1.46$\times10^8$ & $($74.5$\pm$0.8$)\times10^{37}$ & 1.41$\times10^8$ & $($52.7$\pm$0.5$)\times10^{37}$ \\
1994I & 3.46$\times10^8$ & $($1.6$\pm$0.2$)\times10^{37}$ & 3.20$\times10^8$ & $($2.7$\pm$0.3$)\times10^{37}$ & 3.08$\times10^8$ & $($1.9$\pm$0.2$)\times10^{37}$ \\
1996D & 6.45$\times10^6$ & $<$ 5.8$\times10^{39}$ & 5.99$\times10^6$ & $<$ 1.0$\times10^{40}$ & 5.75$\times10^6$ & $<$ 7.1$\times10^{39}$  \\
1996N & 1.64$\times10^7$ & $<$ 5.5$\times10^{37}$ & 1.60$\times10^7$ & $<$ 9.0$\times10^{37}$ & 1.53$\times10^7$ & $<$ 6.5$\times10^{37}$  \\
1996aq & 2.95$\times10^6$ & $<$ 7.7$\times10^{38}$ & 2.71$\times10^6$ & $<$ 1.3$\times10^{39}$ & 2.63$\times10^6$ & $<$ 9.4$\times10^{38}$  \\
1997X & 4.12$\times10^6$ & $<$ 6.3$\times10^{37}$ & 3.75$\times10^6$ & $<$ 1.1$\times10^{38}$ & 3.67$\times10^6$ & $<$ 7.8$\times10^{37}$  \\
1998T & 4.89$\times10^7$ & $($10.1$\pm$0.6$)\times10^{39}$ & 4.55$\times10^7$ & $($1.7$\pm$0.1$)\times10^{40}$ & 4.38$\times10^7$ & $($12.3$\pm$0.8$)\times10^{39}$ \\
1998bo & 5.93$\times10^6$ & $<$ 7.0$\times10^{39}$ & 5.64$\times10^6$ & $<$ 1.2$\times10^{40}$ & 5.41$\times10^6$ & $<$ 8.4$\times10^{39}$  \\
1998bw & 2.09$\times10^7$ & $($2.4$\pm$0.3$)\times10^{39}$ & 1.89$\times10^7$ & $($4.1$\pm$0.5$)\times10^{39}$ & 1.86$\times10^7$ & $($2.9$\pm$0.3$)\times10^{39}$ \\
1999dn & 2.40$\times10^7$ & $<$ 4.7$\times10^{38}$ & 2.19$\times10^7$ & $<$ 8.3$\times10^{38}$ & 2.14$\times10^7$ & $<$ 5.8$\times10^{38}$  \\
1999eh & 7.38$\times10^6$ & $<$ 5.3$\times10^{38}$ & 6.76$\times10^6$ & $<$ 9.2$\times10^{38}$ & 6.56$\times10^6$ & $<$ 6.5$\times10^{38}$  \\
1999ex & 4.83$\times10^6$ & $<$ 2.8$\times10^{39}$ & 4.44$\times10^6$ & $<$ 4.9$\times10^{39}$ & 4.30$\times10^6$ & $<$ 3.5$\times10^{39}$  \\
2000cr & 6.63$\times10^6$ & $<$ 1.1$\times10^{39}$ & 6.09$\times10^6$ & $<$ 2.0$\times10^{39}$ & 5.90$\times10^6$ & $<$ 1.4$\times10^{39}$  \\
2000ds & 2.71$\times10^7$ & $<$ 1.8$\times10^{38}$ & 2.48$\times10^7$ & $<$ 3.1$\times10^{38}$ & 2.41$\times10^7$ & $<$ 2.2$\times10^{38}$  \\
2001ig & 2.06$\times10^7$ & $($1.1$\pm$0.2$)\times10^{38}$ & 1.87$\times10^7$ & $($2.0$\pm$0.3$)\times10^{38}$ & 1.83$\times10^7$ & $($1.4$\pm$0.2$)\times10^{38}$ \\
2003L & 1.30$\times10^7$ & $($1.3$\pm$0.2$)\times10^{40}$ & 1.18$\times10^7$ & $($2.3$\pm$0.4$)\times10^{40}$ & 1.16$\times10^7$ & $($1.6$\pm$0.3$)\times10^{40}$ \\
2003bg & 5.39$\times10^7$ & $($16.9$\pm$0.7$)\times10^{38}$ & 4.92$\times10^7$ & $($3.0$\pm$0.1$)\times10^{39}$ & 4.80$\times10^7$ & $($20.7$\pm$0.8$)\times10^{38}$ \\
2003is & 2.96$\times10^6$ & $<$ 7.0$\times10^{39}$ & 2.86$\times10^6$ & $<$ 1.2$\times10^{40}$ & 2.65$\times10^6$ & $<$ 8.6$\times10^{39}$  \\
2004C & 5.81$\times10^7$ & $($7.7$\pm$0.8$)\times10^{38}$ & 5.32$\times10^7$ & $($1.3$\pm$0.1$)\times10^{39}$ & 5.18$\times10^7$ & $($9.4$\pm$1.0$)\times10^{38}$ \\
2004dk & 3.41$\times10^6$ & $($5.7$^{+ 3.3}_{- 2.3}$ $)\times10^{38}$  & 3.14$\times10^6$ & $($9.9$^{+ 5.8}_{- 4.0}$ $)\times10^{38}$  & 3.04$\times10^6$ & $($7.0$^{+ 4.1}_{- 2.8}$ $)\times10^{38}$  \\
2005U & 4.15$\times10^7$ & $($1.1$\pm$0.3$)\times10^{39}$ & 3.85$\times10^7$ & $($1.9$\pm$0.5$)\times10^{39}$ & 3.70$\times10^7$ & $($1.4$\pm$0.4$)\times10^{39}$  \\
2005at & 1.76$\times10^7$ & $<$ 2.0$\times10^{37}$ & 1.72$\times10^7$ & $<$ 3.3$\times10^{37}$ & 1.65$\times10^7$ & $<$ 2.4$\times10^{37}$  \\
2005cz & 5.86$\times10^6$ & $<$ 5.0$\times10^{38}$ & 5.37$\times10^6$ & $<$ 8.7$\times10^{38}$ & 5.22$\times10^6$ & $<$ 6.1$\times10^{38}$  \\
2006ep & 3.40$\times10^6$ & $<$ 4.4$\times10^{39}$ & 3.12$\times10^6$ & $<$ 7.7$\times10^{39}$ & 3.03$\times10^6$ & $<$ 5.5$\times10^{39}$  \\
2007Y & 1.65$\times10^7$ & $<$ 1.3$\times10^{38}$ & 1.61$\times10^7$ & $<$ 2.1$\times10^{38}$ & 1.54$\times10^7$ & $<$ 1.5$\times10^{38}$  \\
2007gr & 6.78$\times10^6$ & $<$ 8.5$\times10^{37}$ & 6.50$\times10^6$ & $<$ 1.4$\times10^{38}$ & 6.31$\times10^6$ & $<$ 1.0$\times10^{38}$  \\
2007ke & 4.51$\times10^7$ & $<$ 7.5$\times10^{39}$ & 4.32$\times10^7$ & $<$ 1.2$\times10^{40}$ & 4.18$\times10^7$ & $<$ 8.8$\times10^{39}$  \\
2008D & 7.52$\times10^6$ & $($6.5$^{+ 2.8 }_{- 2.1}$ $)\times10^{38}$ & 6.90$\times10^6$ & $($1.1$^{+ 0.5}_{- 0.4}$ $)\times10^{39}$  & 6.69$\times10^6$ & $($8.0$^{+ 3.4 }_{- 2.5}$ $)\times10^{38}$  \\
2008bo & 8.16$\times10^6$ & $<$ 3.5$\times10^{38}$ & 7.50$\times10^6$ & $<$ 6.1$\times10^{38}$ & 7.27$\times10^6$ & $<$ 4.3$\times10^{38}$  \\
2009bb & 4.21$\times10^6$ & $($1.7$^{+ 0.9 }_{- 0.6}$ $)\times10^{39}$  & 3.87$\times10^6$ & $($3.0$^{+ 1.5}_{- 1.1}$ $)\times10^{39}$  & 3.76$\times10^6$ & $($2.1$^{+ 1.1 }_{- 0.8 }$ $)\times10^{39}$ \\
2009jf & 4.26$\times10^6$ & $($1.2$\pm$0.1$)\times10^{40}$ & 3.92$\times10^6$ & $($2.0$\pm$0.2$)\times10^{40}$ & 3.80$\times10^6$ & $($1.4$\pm$0.2$)\times10^{40}$ \\
2009mk & 4.20$\times10^6$ & $<$ 4.7$\times10^{38}$ & 3.87$\times10^6$ & $<$ 8.1$\times10^{38}$ & 3.75$\times10^6$ & $<$ 5.8$\times10^{38}$  \\
2010O & 3.72$\times10^7$ & $($17.6$\pm$0.9$)\times10^{39}$ & 3.46$\times10^7$ & $($3.0$\pm$0.2$)\times10^{40}$ & 3.32$\times10^7$ & $($2.2$\pm$0.1$)\times10^{40}$ \\
2011dh & 3.14$\times10^8$ & $($12.1$\pm$0.4$)\times10^{37}$ & 2.91$\times10^8$ & $($20.1$\pm$0.6$)\times10^{37}$ & 2.80$\times10^8$ & $($14.9$\pm$0.4$)\times10^{37}$ \\
2011ei & 4.12$\times10^6$ & $<$ 1.5$\times10^{39}$ & 3.81$\times10^6$ & $<$ 2.6$\times10^{39}$ & 3.67$\times10^6$ & $<$ 1.8$\times10^{39}$  \\
2012ap & 4.11$\times10^6$ & $<$ 3.6$\times10^{39}$ & 3.80$\times10^6$ & $<$ 6.2$\times10^{39}$ & 3.66$\times10^6$ & $<$ 4.4$\times10^{39}$  \\
2013ak & 4.08$\times10^6$ & $($5.4$\pm$1.0$)\times10^{38}$ & 3.80$\times10^6$ & $($9.2$\pm$1.7$)\times10^{38}$ & 3.64$\times10^6$ & $($6.6$\pm$1.2$)\times10^{38}$ \\
2013ge & 7.66$\times10^6$ & $<$ 1.7$\times10^{38}$ & 7.13$\times10^6$ & $<$ 2.9$\times10^{38}$ & 6.83$\times10^6$ & $<$ 2.1$\times10^{38}$  \\     
2014C & 7.95$\times10^6$ & $($8.0$\pm$0.4$)\times10^{39}$ & 7.48$\times10^6$ & $($13.5$\pm$0.6$)\times10^{39}$ & 7.11$\times10^6$ & $($9.8$\pm$0.4$)\times10^{39}$  \\
          \hline \hline
      \end{tabular}
            \newline
(Note that the luminosities presented here do not accurately reflect the luminosity of supernovae due to chosen spectral models or multi-epoch stacking of observations.)
            \end{minipage}
      \end{table*}

\begin{table*} 
 \centering
 \begin{minipage}{180mm}
 \caption{X-ray luminosities and upper limits of X-ray luminosity (post-explosion, part 2)}\label{resultstable_post2}
  \begin{tabular}{@{}lllllllllllll@{}}

  \hline \hline

 & \multicolumn{2}{@{}c}{(4) Accr. pulsar (high abs.)}  & \multicolumn{2}{@{}c}{(5) Hard-state BH (low abs.)} & \multicolumn{2}{@{}c}{(6) Hard-state BH (high abs.)} \\

Name & Exp. map av. & Luminosity & Exp. map av. & Luminosity & Exp. map av. & Luminosity\\
&  value (cm$^2$s) & (erg s$^{-1}$) &  value (cm$^2$s) & (erg s$^{-1}$) &  value (cm$^2$s) & (erg s$^{-1}$) \\
\hline
1983I & 5.84$\times10^7$ & $<$ 4.7$\times10^{37}$ & 5.38$\times10^7$ & $<$ 2.9$\times10^{37}$ & 5.79$\times10^7$ & $<$ 5.4$\times10^{37}$  \\
1983N & 2.46$\times10^8$ & $<$ 4.1$\times10^{36}$ & 3.14$\times10^8$ & $<$ 1.8$\times10^{36}$ & 2.78$\times10^8$ & $<$ 4.1$\times10^{36}$  \\
1983V & 5.66$\times10^7$ & $<$ 1.2$\times10^{38}$ & 6.13$\times10^7$ & $<$ 6.4$\times10^{37}$ & 5.97$\times10^7$ & $<$ 1.3$\times10^{38}$  \\
1984L & 1.73$\times10^6$ & $<$ 1.4$\times10^{39}$ & 2.20$\times10^6$ & $<$ 6.3$\times10^{38}$ & 1.94$\times10^6$ & $<$ 1.4$\times10^{39}$  \\
1985F & 3.19$\times10^6$ & $<$ 1.5$\times10^{38}$ & 4.08$\times10^6$ & $<$ 6.8$\times10^{37}$ & 3.59$\times10^6$ & $<$ 1.5$\times10^{38}$  \\
1990U & 1.15$\times10^7$ & $<$ 9.1$\times10^{38}$ & 1.46$\times10^7$ & $<$ 4.1$\times10^{38}$ & 1.30$\times10^7$ & $<$ 9.3$\times10^{38}$  \\
1991N & 1.62$\times10^7$ & $($7.6$\pm$2.2$)\times10^{38}$ & 2.11$\times10^7$ & $($3.3$\pm$1.0$)\times10^{38}$ & 1.83$\times10^7$ & $($7.8$\pm$2.2$)\times10^{38}$ \\
1993J & 1.36$\times10^8$ & $($83.0$\pm$0.9$)\times10^{37}$ & 1.49$\times10^8$ & $($43.2$\pm$0.5$)\times10^{37}$ & 1.44$\times10^8$ & $($90.3$\pm$0.9$)\times10^{37}$ \\
1994I & 2.79$\times10^8$ & $($3.2$\pm$0.4$)\times10^{37}$ & 3.52$\times10^8$ & $($1.4$\pm$0.2$)\times10^{37}$ & 3.15$\times10^8$ & $($3.3$\pm$0.4$)\times10^{37}$ \\
1996D & 5.25$\times10^6$ & $<$ 1.2$\times10^{40}$ & 6.51$\times10^6$ & $<$ 5.5$\times10^{39}$ & 5.89$\times10^6$ & $<$ 1.2$\times10^{40}$  \\
1996N & 1.48$\times10^7$ & $<$ 1.0$\times10^{38}$ & 1.60$\times10^7$ & $<$ 5.4$\times10^{37}$ & 1.57$\times10^7$ & $<$ 1.1$\times10^{38}$  \\
1996aq & 2.38$\times10^6$ & $<$ 1.6$\times10^{39}$ & 3.01$\times10^6$ & $<$ 7.1$\times10^{38}$ & 2.67$\times10^6$ & $<$ 1.6$\times10^{39}$  \\
1997X & 3.28$\times10^6$ & $<$ 1.3$\times10^{38}$ & 4.28$\times10^6$ & $<$ 5.8$\times10^{37}$ & 3.70$\times10^6$ & $<$ 1.3$\times10^{38}$  \\
1998T & 4.01$\times10^7$ & $($2.0$\pm$0.1$)\times10^{40}$ & 4.93$\times10^7$ & $($9.4$\pm$0.6$)\times10^{39}$ & 4.47$\times10^7$ & $($2.1$\pm$0.1$)\times10^{40}$ \\
1998bo & 5.08$\times10^6$ & $<$ 1.4$\times10^{40}$ & 5.87$\times10^6$ & $<$ 6.7$\times10^{39}$ & 5.53$\times10^6$ & $<$ 1.4$\times10^{40}$  \\
1998bw & 1.66$\times10^7$ & $($4.9$\pm$0.6$)\times10^{39}$ & 2.18$\times10^7$ & $($2.1$\pm$0.2$)\times10^{39}$ & 1.87$\times10^7$ & $($5.0$\pm$0.6$)\times10^{39}$ \\
1999dn & 1.92$\times10^7$ & $<$ 9.8$\times10^{38}$ & 2.48$\times10^7$ & $<$ 4.3$\times10^{38}$ & 2.16$\times10^7$ & $<$ 1.0$\times10^{39}$  \\
1999eh & 5.90$\times10^6$ & $<$ 1.1$\times10^{39}$ & 7.57$\times10^6$ & $<$ 4.9$\times10^{38}$ & 6.67$\times10^6$ & $<$ 1.1$\times10^{39}$  \\
1999ex & 3.89$\times10^6$ & $<$ 5.8$\times10^{39}$ & 4.95$\times10^6$ & $<$ 2.6$\times10^{39}$ & 4.38$\times10^6$ & $<$ 6.0$\times10^{39}$  \\
2000cr & 5.33$\times10^6$ & $<$ 2.3$\times10^{39}$ & 6.79$\times10^6$ & $<$ 1.0$\times10^{39}$ & 6.00$\times10^6$ & $<$ 2.4$\times10^{39}$  \\
2000ds & 2.17$\times10^7$ & $<$ 3.7$\times10^{38}$ & 2.78$\times10^7$ & $<$ 1.7$\times10^{38}$ & 2.44$\times10^7$ & $<$ 3.8$\times10^{38}$  \\
2001ig & 1.63$\times10^7$ & $($2.4$\pm$0.4$)\times10^{38}$ & 2.14$\times10^7$ & $($1.0$\pm$0.2$)\times10^{38}$ & 1.85$\times10^7$ & $($2.4$\pm$0.4$)\times10^{38}$ \\
2003L & 1.03$\times10^7$ & $($2.7$\pm$0.5$)\times10^{40}$ & 1.35$\times10^7$ & $($1.2$\pm$0.2$)\times10^{40}$ & 1.17$\times10^7$ & $($2.8$\pm$0.5$)\times10^{40}$ \\
2003bg & 4.30$\times10^7$ & $($3.5$\pm$0.1$)\times10^{39}$ & 5.57$\times10^7$ & $($15.4$\pm$0.6$)\times10^{38}$ & 4.85$\times10^7$ & $($3.6$\pm$0.1$)\times10^{39}$ \\
2003is & 2.52$\times10^6$ & $<$ 1.4$\times10^{40}$ & 2.82$\times10^6$ & $<$ 7.0$\times10^{39}$ & 2.78$\times10^6$ & $<$ 1.4$\times10^{40}$  \\
2004C & 4.66$\times10^7 $& $($1.6$\pm$0.2$)\times10^{39}$ & 5.99$\times10^7$ & $($7.0$\pm$0.8$)\times10^{38}$ & 5.25$\times10^7$ & $($1.6$\pm$0.2$)\times10^{39}$ \\
2004dk & 2.75$\times10^6$ & $($1.2$^{+ 0.7}_{- 0.5}$ $)\times10^{39}$  & 3.48$\times10^6$ & $($5.3$^{+ 3.1}_{- 2.1}$ $)\times10^{38}$  & 3.09$\times10^6$ & $($1.2$^{+ 0.7}_{- 0.5}$ $)\times10^{39}$ \\
2005U & 3.38$\times10^7$ & $($2.3$\pm$0.6$)\times10^{39}$ & 4.19$\times10^7$ & $($1.1$\pm$0.3$)\times10^{39}$ & 3.79$\times10^7$ & $($2.3$\pm$0.6$)\times10^{39}$  \\
2005at & 1.59$\times10^7$ & $<$ 3.7$\times10^{37}$ & 1.71$\times10^7$ & $<$ 2.0$\times10^{37}$ & 1.69$\times10^7$ & $<$ 4.0$\times10^{37}$  \\
2005cz & 4.70$\times10^6$ & $<$ 1.0$\times10^{39}$ & 6.02$\times10^6$ & $<$ 4.6$\times10^{38}$ & 5.29$\times10^6$ & $<$ 1.0$\times10^{39}$  \\
2006ep & 2.73$\times10^6$ & $<$ 9.2$\times10^{39}$ & 3.48$\times10^6$ & $<$ 4.1$\times10^{39}$ & 3.08$\times10^6$ & $<$ 9.4$\times10^{39}$  \\
2007Y & 1.48$\times10^7$ & $<$ 2.3$\times10^{38}$ & 1.61$\times10^7$ & $<$ 1.2$\times10^{38}$ & 1.58$\times10^7$ & $<$ 2.5$\times10^{38}$  \\
2007gr & 5.99$\times10^6$ & $<$ 1.6$\times10^{38}$ & 6.73$\times10^6$ & $<$ 8.1$\times10^{37}$ & 6.41$\times10^6$ & $<$ 1.7$\times10^{38}$  \\
2007ke & 3.98$\times10^7$ & $<$ 1.4$\times10^{40}$ & 4.46$\times10^7$ & $<$ 7.1$\times10^{39}$ & 4.26$\times10^7$ & $<$ 1.5$\times10^{40}$  \\
2008D & 6.03$\times10^6$ & $($1.3$^{+ 0.6}_{- 0.4}$ $)\times10^{39}$  & 7.70$\times10^6$ & $($6.0$^{+ 2.6 }_{- 1.9}$ $)\times10^{38}$  & 6.80$\times10^6$ & $($1.4$^{+ 0.6}_{- 0.4}$ $)\times10^{39}$  \\
2008bo & 6.57$\times10^6$ & $<$ 7.2$\times10^{38}$ & 8.35$\times10^6$ & $<$ 3.2$\times10^{38}$ & 7.39$\times10^6$ & $<$ 7.4$\times10^{38}$  \\
2009bb & 3.40$\times10^6$ & $($3.5$^{+ 1.8}_{- 1.3}$ $)\times10^{39}$  & 4.32$\times10^6$ & $($1.6$^{+ 0.8 }_{- 0.6}$ $)\times10^{39}$  & 3.82$\times10^6$ & $($3.6$^{+ 1.9}_{- 1.4}$ $)\times10^{39}$  \\
2009jf & 3.44$\times10^6$ & $($2.4$\pm$0.3$)\times10^{40}$ & 4.36$\times10^6$ & $($1.1$\pm$0.1$)\times10^{40}$ & 3.86$\times10^6$ & $($2.4$\pm$0.3$)\times10^{40}$ \\
2009mk & 3.39$\times10^6$ & $<$ 9.6$\times10^{38}$ & 4.29$\times10^6$ & $<$ 4.3$\times10^{38}$ & 3.82$\times10^6$ & $<$ 9.8$\times10^{38}$  \\
2010O & 3.03$\times10^7$ & $($3.6$\pm$0.2$)\times10^{40}$ & 3.75$\times10^7$ & $($16.5$\pm$0.9$)\times10^{39}$ & 3.40$\times10^7$ & $($3.7$\pm$0.2$)\times10^{40}$ \\
2011dh & 2.55$\times10^8$ & $($24.8$\pm$0.7$)\times10^{37}$ & 3.17$\times10^8$ & $($11.4$\pm$0.3$)\times10^{37}$ & 2.86$\times10^8$ & $($25.3$\pm$0.8$)\times10^{37}$ \\
2011ei & 3.34$\times10^6$ & $<$ 3.0$\times10^{39}$ & 4.18$\times10^6$ & $<$ 1.4$\times10^{39}$ & 3.75$\times10^6$ & $<$ 3.1$\times10^{39}$  \\
2012ap & 3.33$\times10^6$ & $<$ 7.4$\times10^{39}$ & 4.16$\times10^6$ & $<$ 3.4$\times10^{39}$ & 3.74$\times10^6$ & $<$ 7.5$\times10^{39}$  \\
2013ak & 3.33$\times10^6$ & $($1.1$\pm$0.2$)\times10^{39}$ & 4.11$\times10^6$ & $($5.0$\pm$0.9$)\times10^{38}$ & 3.73$\times10^6$ & $($1.1$\pm$0.2$)\times10^{39}$ \\
2013ge & 6.25$\times10^6$ & $<$ 3.5$\times10^{38}$ & 7.68$\times10^6$ & $<$ 1.6$\times10^{38}$ & 7.00$\times10^6$ & $<$ 3.5$\times10^{38}$  \\
2014C & 6.58$\times10^6$ & $($16.0$\pm$0.7$)\times10^{39}$ & 7.86$\times10^6$ & $($7.6$\pm$0.3$)\times10^{39}$ & 7.34$\times10^6$ & $($16.5$\pm$0.7$)\times10^{39}$ \\

\hline \hline
      \end{tabular}
                  \newline
(Note that the luminosities presented here do not accurately reflect the luminosity of supernovae due to chosen spectral models or multi-epoch stacking of observations.)
      \end{minipage}
      \end{table*} 

\subsection{Detected pre-explosion sources} \label{sect_detected}
Although for most pre-explosion targets no source flux above the 3$\sigma$ detection threshold was found, there are three targets that do display a clear signal: SN 2004gt, SN 2009jf and SN 2010O.

\begin{figure}
\centering
\includegraphics[width=0.5\textwidth]{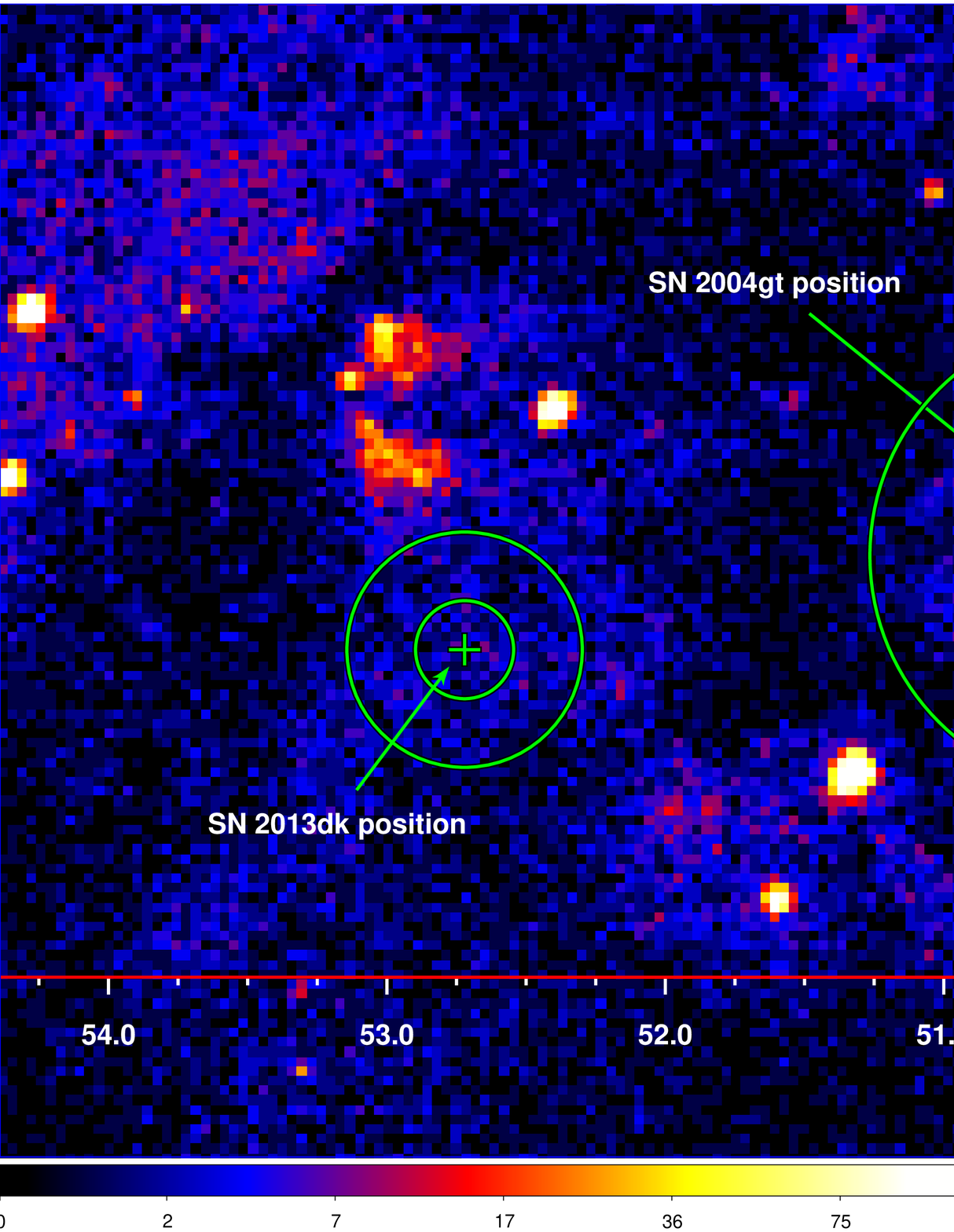}
\includegraphics[width=0.5\textwidth]{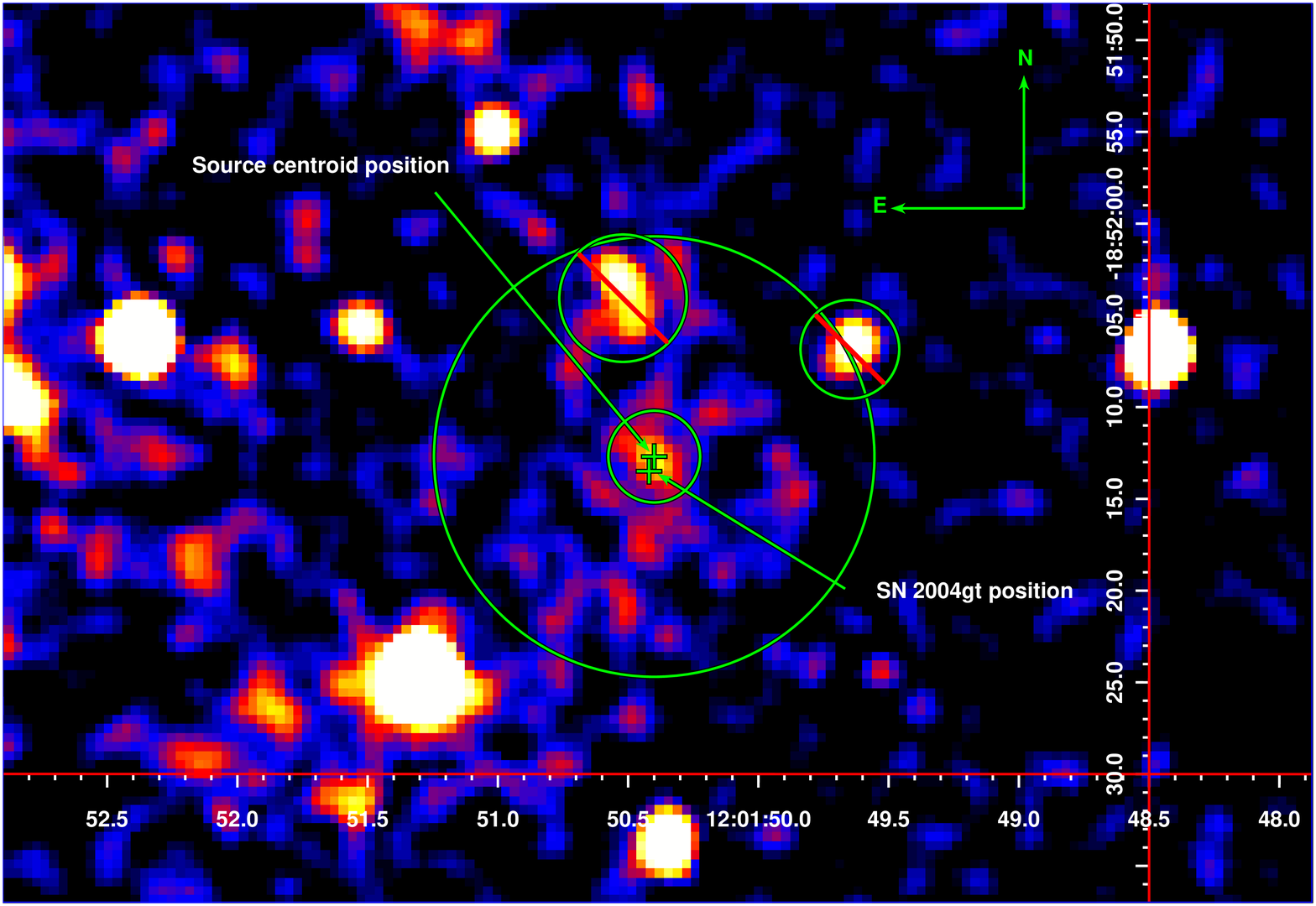}
\caption{Top: SN 2013dk and SN 2004gt (pre-explosion), at the 0.5-7.0 keV energy range. The inner circles are the source apertures, surrounded by the background annuli. The radius of the source aperture is 2.5 arcsec for both cases, and the outer radii of the background annuli are 6 and 12 arcsec, respectively. The apertures are centred on the SN positions. Bottom: The source (CXOU J120150.4-185212) and our measured position of SN 2004gt (pre-explosion) at 1.5-8.0 keV energy range. The central circle is the source aperture (radius 2.5 arcsec) centred on the centroid of the source, and the annulus (outer radius 12 arcsec) around it is the background extraction area. The circles with red bars contain unrelated sources which were excluded from the background extraction area. Each image pixel corresponds to $\sim$0.5 arcsec. FWHM=3-pixel gaussian smoothing has been applied to the bottom image to make the source more visible.}
\label{2013dk_fig}
\label{2004gt_fig}
\label{2004gt_source_fig}
\end{figure}

In the case of SN 2004gt, a complex emission area permeates the entire source region, confusing any signal from a potential progenitor (see Fig. \ref{2004gt_fig}). Because this emission is primarily from low-energy photons, we re-examined this dataset using a higher, 1.5-8.0 keV energy range. First, we corrected the aspect solution of the seven original \textit{Chandra} observations using CIAO's \textit{reproject\_aspect} tool by matching sources found with \textit{wavdetect} with the USNO A2.0 catalog \citep{USNO}, then reprocessing the data the same way as discussed before. We then remade the photon count and exposure maps for SN 2004gt position using the higher 1.5-8.0 keV energy range. The corrected count map was convolved with a Gaussian kernel with $\sigma$ = 1.27 pixels (FWHM = 1.5 arcsec) to enhance the detection of faint point sources. In the convolved image a point-like source becomes visible close to the position of SN 2004gt. We identify this source as the known X-ray source CXOU J120150.4-185212 \citep{2004gt_catalog_source}. To determine whether this pre-explosion source is coincident with the supernova position we determined its centre coordinates with 1-dimensional Gaussian fitting using the IRAF XIMTOOL package and with centroiding and Gaussian fitting as implemented in the IRAF APPHOT package. These three methods gave similar coordinates with a standard deviation of 0.4 pixels. However, the uncertainty in the centre coordinates is probably larger than this given the low S/N of the object in the original unconvolved data and we therefore adopt 1 pixel (corresponding to 0.5 arcsec) as the uncertainty in its (x,y) position. Combining this error with the uncertainty in the absolute astrometric calibration yields a total error of $\pm$0.6 arcsec for both the RA and Dec of the pre-explosion source (at the distance of SN 2004gt, 1 arcsec corresponds to a projected distance of $\sim$100 pc). We therefore adopt the position RA = 12:01:50.40, Dec = -18:52:12.7 for the source. 

We also derived new absolute astrometry for SN 2004gt making use of near-infrared Ks-band imaging of NGC 4038/9 obtained as a part of a search for SNe in the nuclear regions of starburst galaxies \citep*[see][]{mattila_starburst} with the William Herschel Telescope (WHT) LIRIS instrument on 2005 January 30 and INGRID instrument on 2002 January 3. The images were reduced using standard procedures in IRAF \citep[see][]{kangas_2013}. For this the World Coordinate System (WCS) of the LIRIS image was calibrated making use of the centroid positions of 26 point-like sources (incl. both foreground stars and star clusters in NGC 4038/9) together with their coordinates from the 2MASS catalog. The SN is located in proximity of a bright star cluster making the precise determination of its location in ground-based seeing limited images more difficult. To obtain accurate coordinates for the SN we therefore aligned the pre-explosion INGRID image to the LIRIS image and performed image subtraction using the ISIS 2.2 package \citep[][]{alard_1998, alard_2000}. The centroid position of the SN measured from the subtracted image corresponds to RA = 12:01:50.42, Dec = -18:52:13.5, with uncertainty of 0.16 arcsec in RA and 0.19 arcsec in Dec dominated by the uncertainty in the WCS transformation. The SN is therefore coincident with the derived position of the source within $\sim1\sigma$.

We measured the flux of this source in the higher energy range, and found 50 photons inside a source aperture with a 2.5 arcsec radius. A scaled background of 25 photons was obtained from an annulus with outer radius of 12 arcsec centred on the source (excluding two interfering sources from the background, see Fig. \ref{2004gt_source_fig}). The number of photons was too low to obtain a spectrum, and we therefore measured the luminosity using the same models we used for the other SNe\footnote{The average photon energies and absorption correction factors were scaled accordingly for the 1.5-8.0 keV energy range, with other parameters being the same as in Table \ref{models}.}. Rescaling from 1.5-8.0 keV to the 0.5-7.0 keV energy range, we obtained the unabsorbed luminosity of $(3.7\pm1.1)\times10^{37}$ erg s$^{-1}$ for the low-absorption 1 keV temperature blackbody (model 1), $(5.8\pm1.6)\times10^{37}$ erg s$^{-1}$ for the high-absorption 0.5 photon-index powerlaw (model 4), with the luminosities for other models falling between these two values. The average luminosity from all models is $4.6\times10^{37}$ erg s$^{-1}$. We note that this constitutes a detection with an approximately 3$\sigma$ significance. Note that the values presented in Table \ref{resultstable_pre1} are the measurements for the original 0.5-7.0 keV energy range at the SN position using the extraction-regions depicted in Fig. \ref{2004gt_fig}. Because the emission in this energy range most likely originates from the diffuse emission field permeating the region, these measurements do not accurately reflect the nature of the source discussed above, although it is contained within the source aperture. We also note the proximity of a bright star cluster approximately 1 arcsec North and 0.6 arcsec West from the SN position \citep[see Fig. 1 in][]{maund_2005}, possibly contributing some X-ray sources/emission to the vicinity of the SN location \citep[see eg.][for possible associations between star clusters and HMXBs]{juri_ulx}.

In the case of SN 2009jf the flux is from the ultraluminous X-ray source CXOU J230453.0+121959, which overlaps with the SN position and could potentially be associated with the SN, although this cannot be confirmed \citep{ref26}. The post-explosion observation, taken $\sim$ 30 days after the SN explosion (2009 September 23, from \citet{ref10}), shows the ULX with a substantially increased luminosity ($L_\mathrm{diff} \sim 4.3\times10^{39}$ erg s$^{-1}$). As \citet{ref26} also note, this would be consistent with the increase in the flux being caused by the SN.

\begin{figure}
\centering
\includegraphics[width=0.5\textwidth]{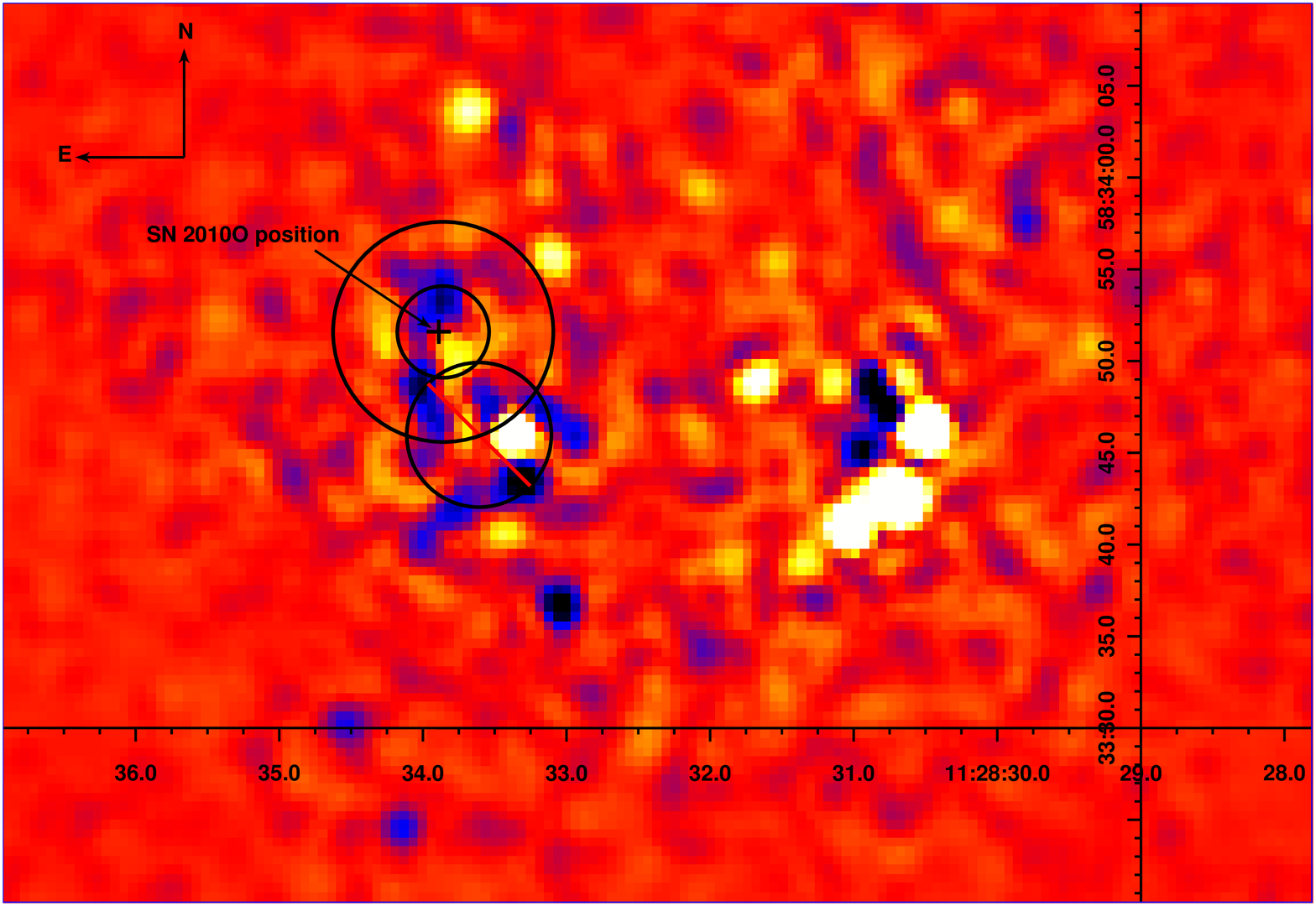}
\caption{Exposure-corrected difference-map of SN 2010O position in Arp 299 (cm$^{-2}$s$^{-1}$) (Spectral model: 1 keV blackbody, $N_\mathrm{H}=0.5\times10^{22}$ cm$^{-2}$, darker areas indicate reduction in flux in the post-explosion data). FWHM=3-pixel gaussian smoothing has been applied to the image. The concentric circles depict the source aperture and the background region used for extraction. The circle with a red bar contains numerous unrelated bright sources which were excluded from the extraction area. The cross marks the position of the transient source found by \citet{ref2}.}
\label{2010O_diffmap}
\end{figure}

SN 2010O in the galaxy Arp 299 has an extremely complex background, which makes any measurements taken for this source difficult. \citet{ref2} suggested the detection of a progenitor for SN 2010O by measuring the X-ray flux variability between the two \textit{Chandra} pre-explosion datasets (OBSIDs 1641 and 6227). We had, in addition to these, two more recent post-explosion datasets (OBSIDs 15077 and 15619). We subtracted the combined (exposure-corrected) pre- and post-explosion observations from each other to see how much variability there was in the source aperture area (see Fig. \ref{2010O_diffmap} for the exposure-corrected difference map). While there is variability within the source aperture, it is widely distributed spatially and could not be attributed to any single point source (notably, SN 2010O cannot be clearly identified from the post-explosion dataset). This suggests the area contains multiple overlapping variable sources within a very complex X-ray background, and identifying any particular sources inside the source aperture could not be made with any confidence. We have included the measured luminosities for this SN in our sample for completeness but note that the results presented here cannot be associated with a progenitor but rather reflect the complex nature of the extraction area. The post-explosion measurements of SN 2010O likewise reflect the complexity of the region, rather than any clear source that could be associated with the SN.

\subsection{Detected post-explosion sources} \label{sect_detected_post}
\begin{figure}
\centering
\includegraphics[width=0.5\textwidth]{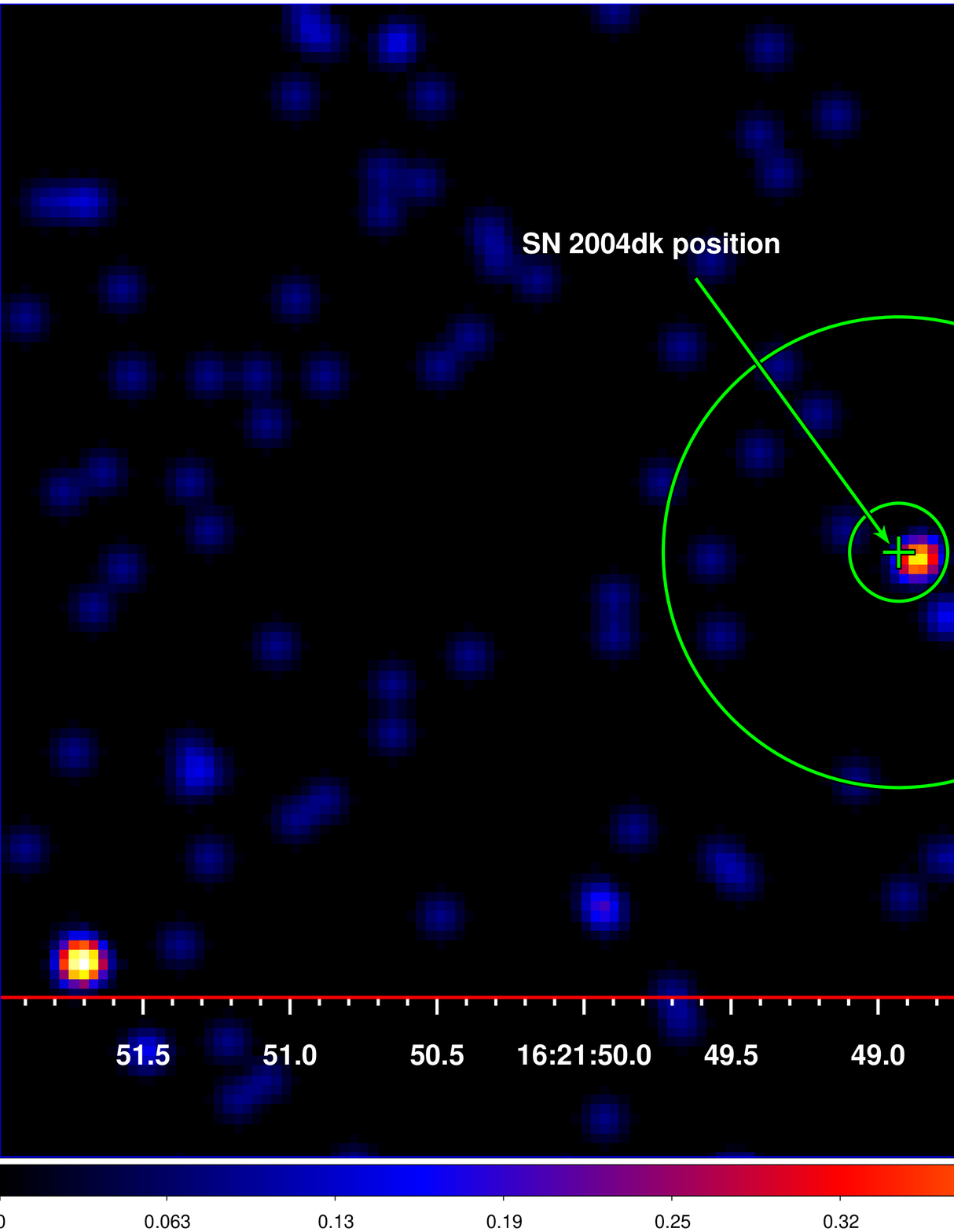}
\includegraphics[width=0.5\textwidth]{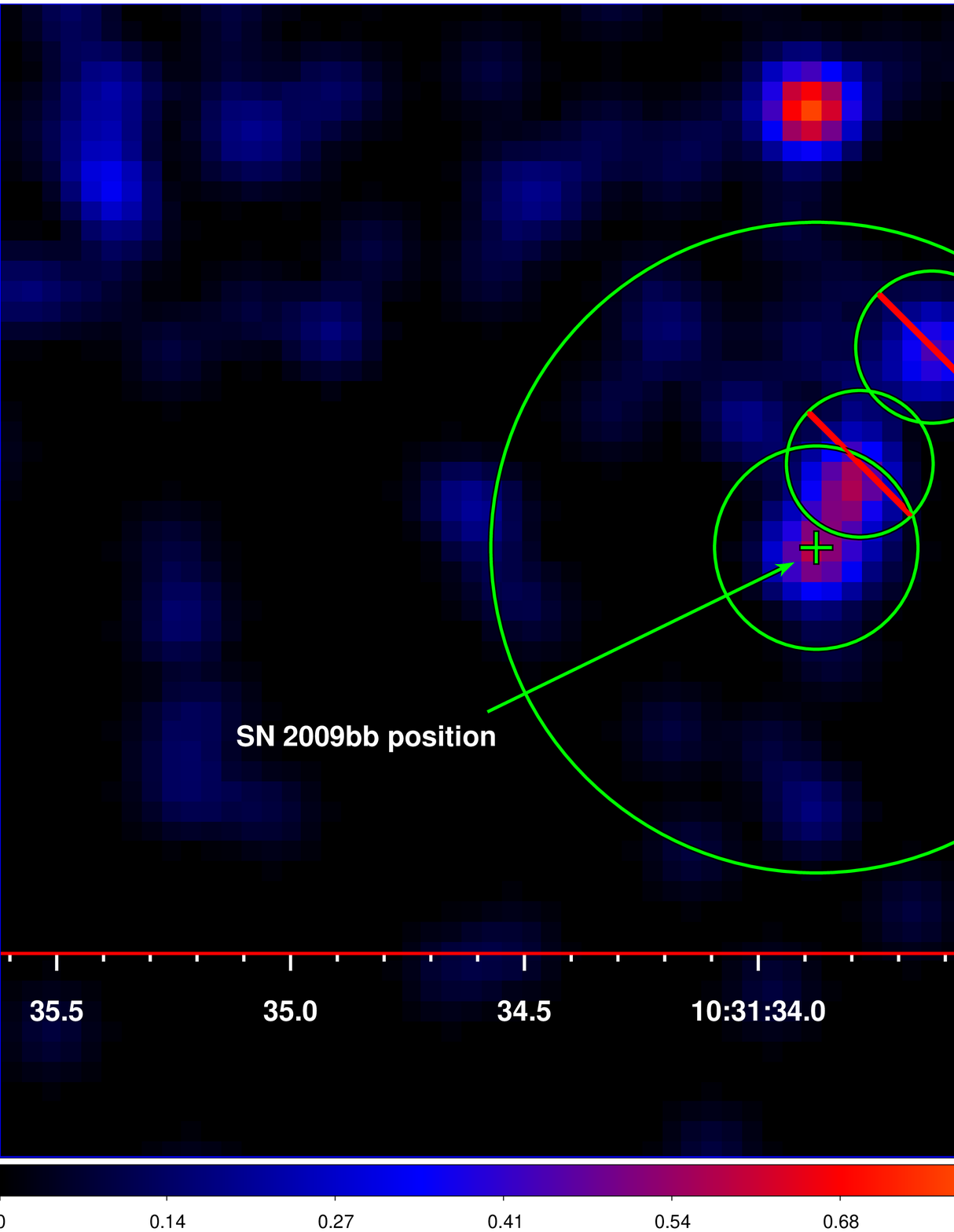}
\includegraphics[width=0.5\textwidth]{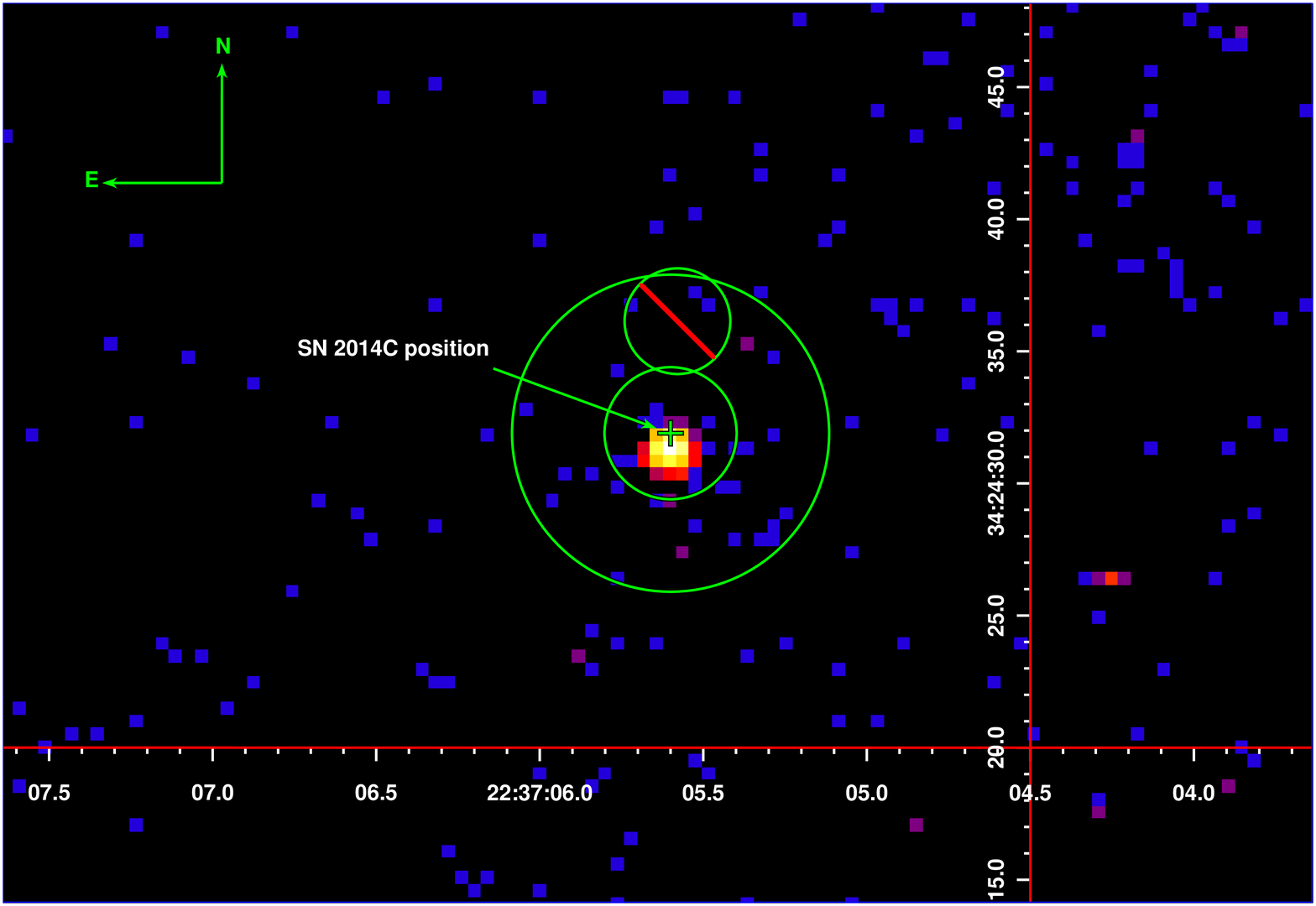}
\caption{The post-explosion sources at the positions of SN 2004dk, SN 2009bb and SN 2014C. FWHM=3-pixel gaussian smoothing has been applied to the images of SN 2004 dk and SN 2009bb to make the sources more visible. The circles with red bars contains sources which were excluded from the extraction areas. The exclusion-circle near SN 2014C is at the position of a variable source visible in the pre-explosion data. Note that the source at SN 2009bb position partially blends with a nearby excluded source.}
\label{unconfirmed_SNe}
\end{figure}

\begin{figure}
\centering
\includegraphics[width=0.5\textwidth]{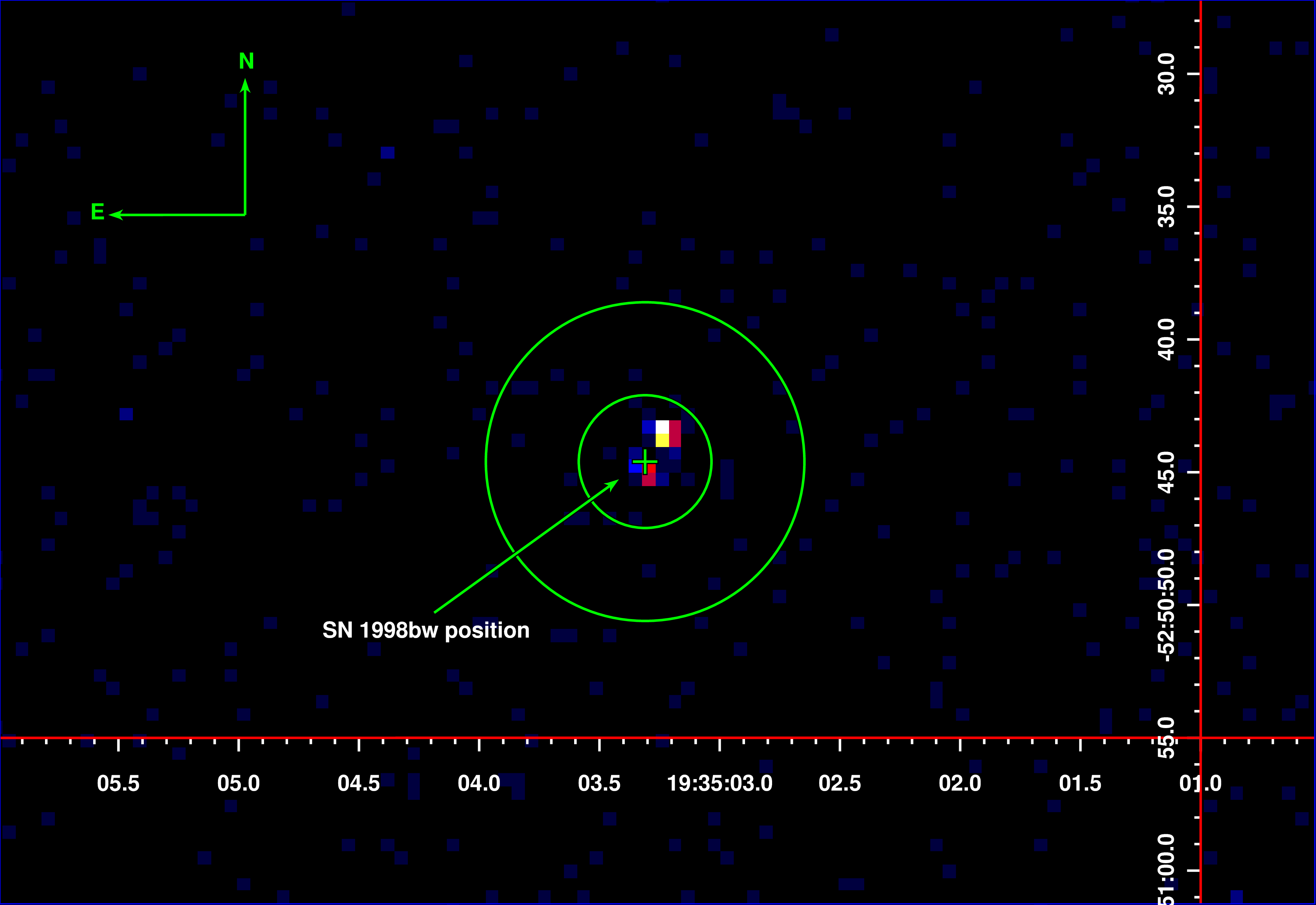}
\includegraphics[width=0.5\textwidth]{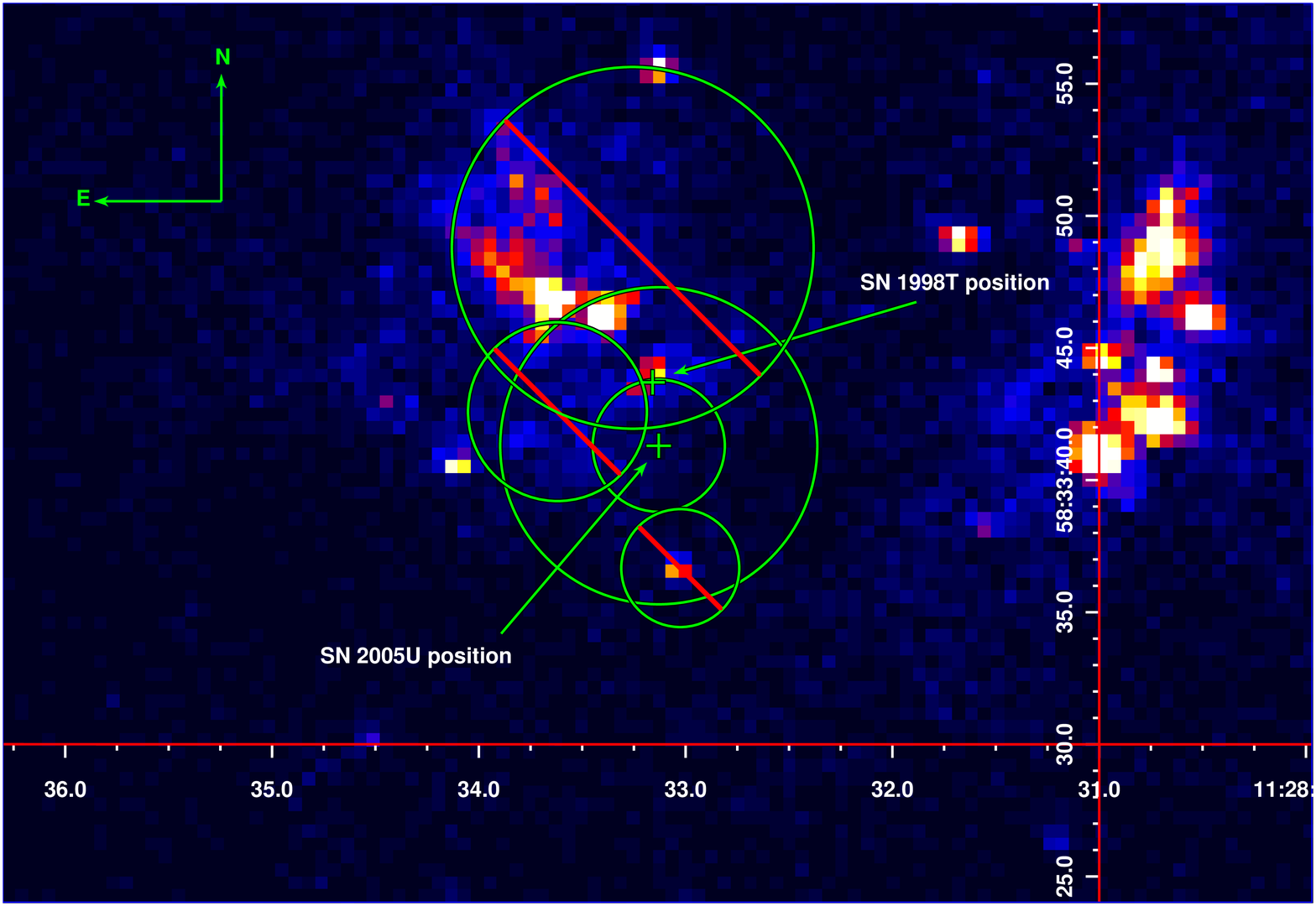}
\caption{Top: SN 1998bw position (post-explosion). Bottom: SN 2005U position (post-explosion). The central circle is the source extraction area and the annulus around it is the background extraction area. The circles with red bars contain unrelated sources which were excluded from the extraction area. The relative location of SN 1998T is also marked.}
\label{1998bw_fig}
\label{2005U_fig}
\end{figure}

In the post-explosion observations, in addition to SN 2009jf and SN 2010O, there are 16 other sources that have flux above the 3$\sigma$ detection threshold. Of these, in the cases of SN 1993J \citep{ref30}, SN 1998bw \citep{ref28}, SN 2001ig \citep{ref31}, SN 2003L \citep{dist28_b}, 2003bg \citep{2003bg_disc}, SN 2004C \citep{ref34}, SN 2008D \citep{ref6}, SN 2011dh \citep{ref32} and SN 2013ak \citep{ref33} the source is the supernova itself. A notable example is the case of SN 2008D, where the observation was taken only 10 days after the explosion, which is well recorded as it was serendipitously discovered by \citet{ref6} while observing a different SN in the same galaxy using the \textit{Swift} X-ray telescope. The sources at the positions of SN 2004dk, SN 2009bb and 2014C (see fig. \ref{unconfirmed_SNe} are most likely the SNe themselves as well given the timing of the observations, though we could not confirm these from literature. Note that SN 2009bb is partially blended with a nearby source. Because characterising the properties of these sources is outside the scope of this paper, we did not attempt to disentangle them. Similarly, in the case of SN 1998bw, the source aperture actually contains two sources at close proximity (Fig. \ref{1998bw_fig}), of which the source closer to the centre of the source aperture is SN 1998bw \citep[for details, see][]{ref28}. Again, because the sources partially overlap, we did not attempt to disentangle them. 

SN 1998T and SN 2005U are in the same galaxy as SN 2010O, Arp 299, and in close enough proximity to each other ($\sim$2.5 arcsec) that their source apertures partially overlap. Similar to SN 2010O, the region is very complex and it is difficult to identify particularly faint sources from the area. The position of SN 1998T overlaps with a known ULX, [ZWM2003] 12 \citep*{2005U_zezas}, which we identify as the source. SN 2005U had both pre- and post-explosion data available, and for both measurements we excluded the ULX and other interfering sources (see Fig. \ref{2005U_fig}) from both the background and the source aperture. Although this produces an apparent detection at the position of SN 2005U post-explosion (the flux from within the defined source aperture exceeds the background by 3$\sigma$), we associate this with the inhomogeneous distribution of diffuse emission in the region, rather than any point-sources within the aperture (the reason why this signal is not apparent in the pre-explosion observation is likely due to the shorter exposure time -- the luminosity of this apparent source is lower than the upper limit established for the pre-explosion case). The measured luminosities presented in tables \ref{resultstable_post1} and \ref{resultstable_post2} for this SN should therefore be mainly treated as upper limits.

In the case of SN 1994I we see three faint point sources inside the source aperture, one of which appears to be very close to the SN position, but overall all three are relatively faint and within \textit{Chandra's} 90 percent encircled energy radius for the PSF which makes associating any of them with a particular source within this source aperture unreliable, although the central one most likely is the SN \citep*[also see][]{ref36}. For the purposes of this study and unlike in the case of SN 1998bw, because the sources do not overlap as clearly, we have excluded the other sources and measured the central source, assuming it to be the SN. Similarly, with SN 1991N, the source region is extremely crowded, with high background emission and multiple nearby bright point sources. There are also several possible sources inside the source aperture which cannot be directly associated with the SN position or identified with confidence to exclude, which together produce enough flux to rate as a false detection. 

Finally, in the case of SN 1983I, we note that there is some uncertainty to the coordinates of this supernova (Asiago catalog only reports the coordinates of the host galaxy while USC and SSC both report different coordinates $\sim$2 arcsec apart\footnote{Both calculated from the same nuclear offsets, possibly based on \citet{1983I_disc}, which may have a high uncertainty.}). The coordinates used in this study are those reported by USC. There is a clear X-ray point-source $\sim$3 arcsec from the USC position which we were unable to conclusively identify and therefore cannot rule out that this is actually SN 1983I. Note that the luminosities for post-explosion sources presented in this paper are not directly representative of the SNe, as our models represent HMXBs and were not fitted to the individual spectra of the SNe. Similarly, some of the measurements (such as those of SN 1993J) involved stacking observations across multiple epochs over a potentially long time-period, and therefore would include any variability in the source during that timeframe.

\subsection{SN luminosity upper limits} \label{sect_ulim}
\begin{figure*}
\centering
\includegraphics[width=0.9\textwidth]{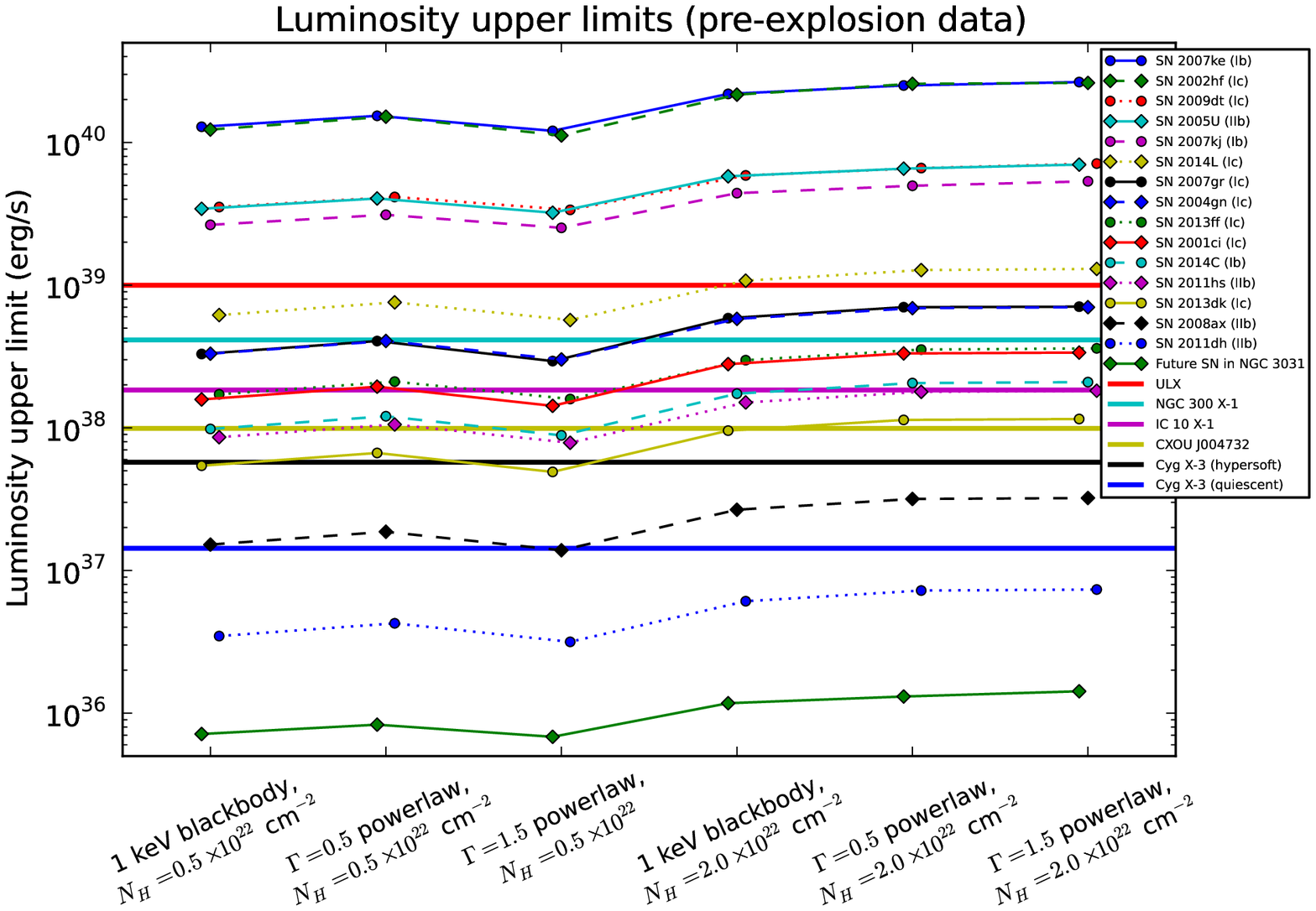}
\caption{Pre-explosion luminosity upper limits. SNe with a detected source are not included (for details, see section \ref{sect_detected}).}
\label{pre_sn_plot}
\end{figure*}

\begin{figure*}
\centering
\includegraphics[width=0.9\textwidth]{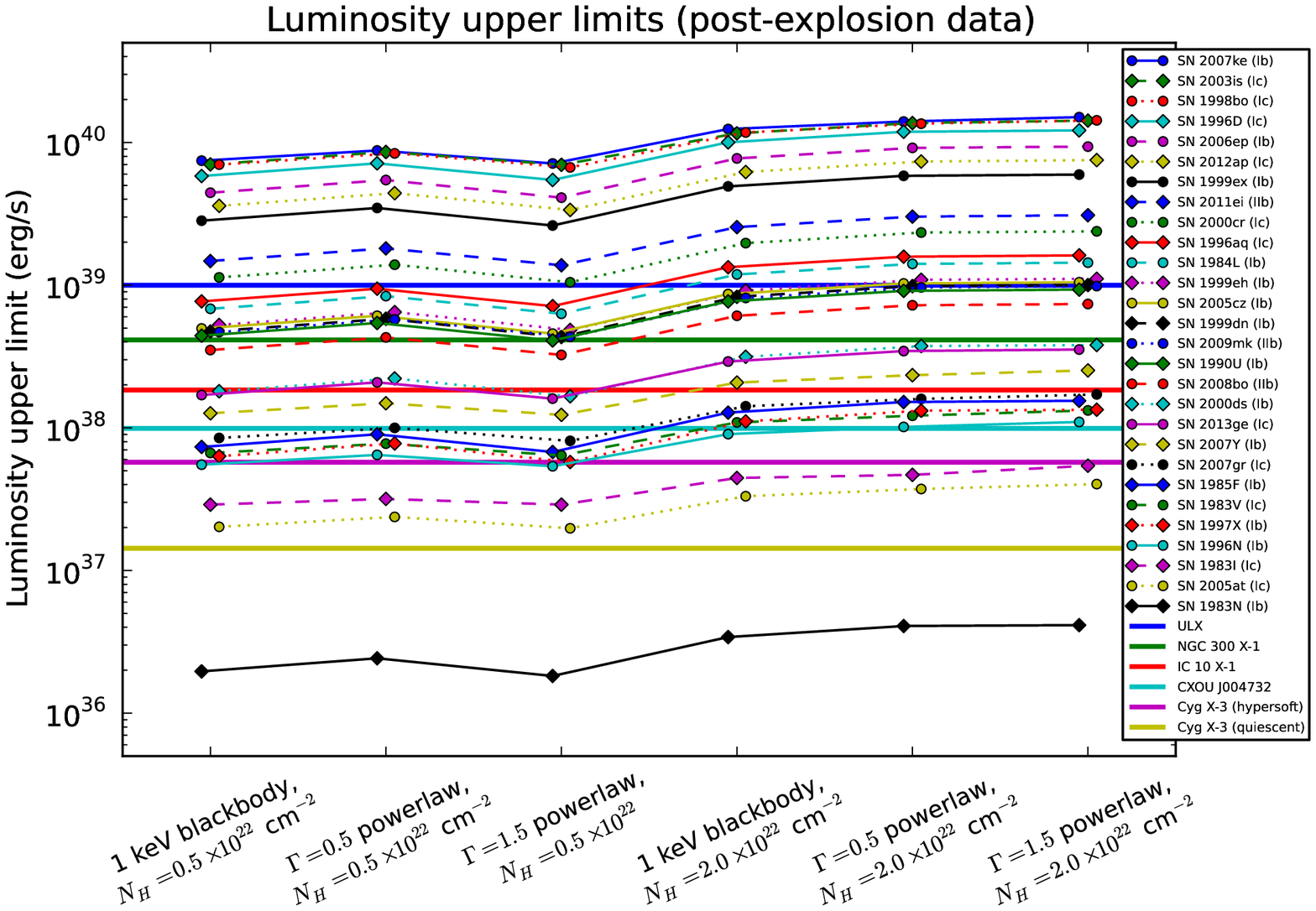}
\caption{Post-explosion luminosity upper limits. SNe with detected source are not included (for details, see section \ref{sect_detected}).}
\label{post_sn_plot}
\end{figure*}

Figures \ref{pre_sn_plot} and \ref{post_sn_plot} show the measured upper limits plotted for the six spectral models used. Targets with a detection above 3$\sigma$ are not included. Because Wolf-Rayet stars as a part of HMXB systems will result in stripped-envelope SNe, in both plots we also compare our results against the luminosities of the three Wolf-Rayet X-ray binaries that are known to exist. These are NGC 300 X-1 with average unabsorbed luminosity $4.1\times10^{38}$ erg s$^{-1}$ \citep{ref12}, IC 10 X-1 with unabsorbed luminosity $1.8\times10^{38}$ erg s$^{-1}$ \citep{ref13} and Cygnus X-3, presented in two modes: quiescent with $L=1.4\times10^{37}$ erg s$^{-1}$ and hypersoft with $L=5.7\times10^{37}$ erg s$^{-1}$ \citep*[both from][representing the states with the lowest and highest X-ray fluxes, respectively]{ref14}. A fourth candidate WR-binary CXOU J004732.0-251722.1 in the galaxy NGC 253 identified by \citet{ref11} is also included (with unabsorbed luminosity $L=9.9\times10^{37}$ erg s$^{-1}$). The aforementioned luminosities have been re-scaled to our 0.5-7.0 keV energy range from their original values. Finally, we also include a comparison to ultraluminous X-ray sources (ULX), which are defined as sources with X-ray luminosity above $1\times10^{39}$ erg s$^{-1}$.

\subsection{Fraction of X-ray bright progenitor systems} \label{sect_eldridge}

There are two requirements for a progenitor to be a luminous X-ray source in a pre-explosion observation. The first is for the progenitor star to have a compact object as a binary companion. The second is that the binary separation must be small enough for the progenitor to supply mass to the companion, either by Roche-lobe overflow or stellar wind, to produce the X-ray luminosity.

We have attempted to gain an estimate of the number of binary systems that might be luminous in X-rays when the system explodes. We have used the latest version (2.0) of the Binary Population and Spectral Synthesis, BPASS, stellar population models \citep*[][Eldridge et al., in prep; http://bpass.auckland.ac.nz]{new_bpass, bpass}. The stellar population included stars at solar metallicity with initial masses from 0.1 to 300M$_{\odot}$. We determined the mass of the primary using a Salpeter initial mass function (IMF) with a slope of $-$2.35. We then use a binary period distribution that is flat in log of the period and flat in mass ratio distribution to determine the mass of the secondary. These are consistent with the distributions observed by \citet{sana}. The orbits are assumed to be circular. While observed binaries have eccentric orbits, we note that if the stars are close enough to interact, circularization will most likely have occurred and, as found by \citet{hurley}, a full treatment of eccentricity is not generally required in population synthesis.

We assume that a star explodes if it has a final mass greater than 2M$_{\odot}$ and a CO core mass greater than 1.38M$_{\odot}$. We count the stellar models which will explode as type Ib, Ic or IIb SNe being those which have less than 0.5M$_{\odot}$ hydrogen at the point of core-collapse. We then also record each event that occurs within a binary with a compact object where the size of the progenitor model is comparable to the size of the orbit and therefore is likely to be emitting X-rays via either Roche-lobe overflow or by wind accretion. This is a simple way to estimate the number of progenitors that might be in X-ray binary systems. We also separately record SNe that are the first ones to occur in the binary system, although these are less likely to be detectable in pre-explosion X-ray observations, as wind-wind collision systems have much weaker X-ray emission than HMXBs.

We show these estimates in Table \ref{eldridgetable}, where the first column is the lower limit for the ratio of the SN progenitor's radius to the binary separation. The second column shows the percentage of all stripped-envelope SNe (in binary systems) that have the corresponding or higher ratio of progenitor radius to binary separation. The third column is the same as the second one, except only for systems where the progenitor's companion has already exploded as a SN, producing a compact object. From this, we can see that the fraction of stripped envelope progenitor systems that are likely to be luminous X-ray sources is only up to a few percent. This number could be significantly higher if close binaries without a compact object, where stellar winds interact and produce X-rays (with significantly lower luminosities), were also included. However, the values provided by our modelling are uncertain and depend on many factors. They could be increased if, for example, mass-transfer is more efficient than assumed or supernova kicks are weaker than assumed. We therefore stress that these values are only rough estimates, but they do at least indicate the order of magnitude for the expected rate.

\begin{table}
 \centering
  \begin{minipage}{140mm}
  \caption{Stripped envelope SN progenitors as a function \hspace{\textwidth}mm of binary separation.}\label{eldridgetable}
  \begin{tabular}{@{}llllllllll@{}}
  \hline
Ratio of progenitor & Fraction of binary & Fraction of binary  \\
 radius to separation, & systems with a & systems with a  \\
 $(R_* / a)$ & normal companion &  compact companion\\
\hline  
$>$ 1      &     0.07  & 0.004 \\
$>$ 0.5    &     0.18  & 0.009 \\
$>$ 0.1    &     0.39  & 0.018 \\
$>$ 0.01   &     0.72  & 0.026 \\
   \hline
 \end{tabular}
 \end{minipage}
\end{table}

\subsection{Upper limits for future nearby SNe} \label{sect_future_SN}
In order to investigate the potential of the presented method for constraining the progenitor system of a future SN in a nearby galaxy, we measured the upper limit of unabsorbed luminosity for an empty field in the host galaxy of SN 1993J, NGC 3031, 10.2 arcsec northeast of the SN and using the same set of observations. We chose this galaxy not only because it is the host to the closest SN in our sample (at distance 3.63 Mpc), but also because it has one of the longest combined exposure times available (752.3 ks). The extraction area was a circular aperture with 2.5 arcsec radius, which was found to contain 24 photons for the combined observation. For the spectral models used in this study, the range of upper limits of luminosity (corrected for absorption) obtained were $L_\mathrm{UL}=(0.7-1.4)\times10^{36}$ erg s$^{-1}$ for models 5 and 6 respectively, with other models falling between these values. This gives us a reasonable estimate for the faintest progenitor source that could be detected for a future extragalactic SN. It is notable that this is almost an order of magnitude fainter than the quiescent state of Cygnus X-3, the closest and faintest known WR X-ray binary at the distance of $\sim$9 kpc \citep{ref14}.

\section{Discussion} \label{sect_discussion}
Of the 18 SNe with pre-explosion data available, three were found to have flux above the 3$\sigma$ detection threshold at their positions. Two of these are identifiable point sources and possible candidates for a HMXB SN progenitor, both previously known sources. The possible association between the ULX CXOU J230453.0+121959 and SN 2009jf was already suggested by \citet{ref26}. We also consider it possible that CXOU J120150.4-185212 might be associated with the progenitor of SN 2004gt. \citet*{mineo} categorized this source as a HMXB based on its relative position within the host galaxy, and although we note that this categorization is purely statistical and has not been confirmed observationally, it does suggest that the source can plausibly be considered a HMXB progenitor candidate for the SN. The lack of post-explosion data for SN 2004gt prevents the determination of any variability in this source, however. Given that over a decade has passed since the explosion of SN 2004gt, the SN may have faded enough that it might be possible in the near-future to measure this source again for possible variability, although given the relative faintness of the source in the pre-explosion data (with luminosity $\sim5\times10^{37}$ erg s$^{-1}$) and the overall complexity of the region, this may be difficult. The third detection is at the location of SN 2010O, which contains a complex system of multiple variable sources. There is a possibility that SN 2010O has an X-ray bright progenitor as suggested by \citet{ref2}, although as noted previously, the complexity of the region makes associating any particular source with the SN difficult. All three of the aforementioned potential associations between X-ray sources and their corresponding supernovae remain uncertain, however.

One SN in our sample, SN 2008ax is known to not have a HMXB progenitor from optical pre-explosion data. However, for the 14 other SNe, X-ray luminosity upper limits were established, constraining the properties of any progenitor systems. Although all possible HMXB progenitor systems cannot be ruled out based on these limits, some of the most luminous cases can be excluded. Specifically, a ULX progenitor could be ruled out for 9 out of 14 SNe in the sample. For one of these SNe, SN 2011dh, optical pre-explosion observations had previously identified a yellow supergiant progenitor in a binary system. However, the recent observations by \cite{maund_2015} suggest that the companion to the progenitor is either fainter than previously believed, or may not exist at all. As the nature of the companion, if one exists, is uncertain, SN 2011dh makes an interesting target for our study. The relatively low upper limit of X-ray luminosity evident from our results could be taken as an indication that this SN did not have a HMXB progenitor.

We also compared the upper limits to the luminosities of the three known (and one candidate) Wolf-Rayet X-ray binaries, a type of HMXB likely to produce a stripped-envelope CCSN. The known Wolf-Rayet HMXBs tend to be more X-ray luminous than HMXBs on average: NGC 300 X-1, IC 10 X-1 and CXOU J004732 all have X-ray luminosities that are higher than $10^{38}$ erg s$^{-1}$, whereas the majority of HMXBs appear to be less luminous \citep[see e.g.][]{mineo}. However, the case of Cygnus X-3, the only known WR X-ray binary in our own galaxy, demonstrates both that much fainter systems of this nature can exist and that their luminosity can be greatly variable depending on the state of the system. In general, given the relative rarity of WR-HMXBs, it is quite likely that there is an observational bias favoring the detection of only these brighter systems. The upper limits obtained in this study are mostly comparable to the average luminosity of the known WR-HMXBs, suggesting that such systems could in principle have been detected.

One problem inherent in interpreting pre-explosion X-ray observations is that it is not known whether or not accretion from the progenitor to the compact companion continues all the way to the explosion, and if not, how long before the core-collapse it would shut down. Because this accretion is the source of X-ray flux from the assumed progenitor system, it is possible that the system has ceased to be a luminous X-ray source before the SN on a timescale which would make pre-explosion detection impractical or impossible. Our BPASS-modelling discussed earlier indicates that, while a significant number of stripped-envelope SN progenitors can be expected to be found in binary systems, only a small fraction of these would contain a compact companion at a binary separation where substantial mass-transfer could be expected to occur. HMXB-type progenitors would therefore be quite rare. The possibility of observing X-rays from non-HMXB binary progenitors remains, but such objects would most likely be relatively faint (comparable to a WR-star), and well below the threshold found in this study.

One way to constrain the timeframe in which a progenitor system might become X-ray bright is to observe HMXBs that are associated with supernova remnants (SNR). SNRs are relatively short-lived, so this would help establish how soon after the supernova accretion might be initiated in a newly formed HMXB. Several such systems have been identified, for example SXP 1062 \citep{SNR_brunet} and DEM L241 \citep{SNR_seward}, both of which have an estimated age of several times $10^{4}$ years. Another possible case is Circinus X-1, for which \citet{SNR_heinz} derive an age upper-limit of $\sim 4600$ years. While none of these systems are X-ray luminous enough to allow detection by our study, these examples indicate that accretion in a newly-formed HMXB might begin on a timescale that is less than a few thousand years after the SN. Another example to consider is the association between SNR W50 and SS 433, although it is estimated to be older than the aforementioned cases \citep*[see eg.][]{SNR_goodall}.

X-ray binaries are not the only type of pre-explosion X-ray source that might be associated with a SN. Colliding wind binaries (CWB), such as the luminous blue variable (LBV) binary Eta Carinae, or the Wolf-Rayet binary WR 48a, are known to have X-ray luminosities up to $\sim 10^{35}$ erg s$^{-1}$ \citep*[see e.g.][]{etacar, lbv_sur, zhekov}. Compared to the upper limits we derived, such objects are an order of magnitude fainter than what could reasonably be detected with \textit{Chandra}, unless they were significantly closer. Also, given that eg. Eta Carinae is known to have a fairly hard X-ray spectrum \citep[eg.][]{hamaguchi}, they would likely be difficult to distinguish from HMXBs based on their spectrum alone. Future studies are therefore unlikely to find such SN progenitors using this method.

\section{Conclusions} \label{sect_conclusions}
The aim of this study was to examine the possibility of high-mass X-ray binaries being the progenitors to some stripped-envelope CCSNe with a direct observational test, or to constrain the possibility of such progenitors in the sample. If a SN progenitor was part of a HMXB before the explosion, the system could be an X-ray source due to accretion from the progenitor star to the compact object. In this study we note the existence of two pre-explosion sources that could potentially be associated with supernovae. The first of these is the ULX CXOU J230453.0+121959 at the position of the type Ib SN 2009jf \citep{ref26}, with luminosity $\sim1\times10^{40}$ erg s$^{-1}$. The second candidate is CXOU J120150.4-185212, at the position of the type Ic SN 2004gt, with luminosity $\sim5\times10^{37}$ erg s$^{-1}$. For both of these, however, it is important to note that chance co-alignment is possible. We also measure the unabsorbed pre-explosion upper limits of X-ray luminosity for the positions of 14 stripped-envelope CCSNe. Most of the upper limits obtained are comparable to the luminosities of the two known and one candidate extragalactic WR-HMXBs, showing that at least more luminous examples of such systems could in principle have been detected. Therefore, while detecting a candidate progenitor in this way is possible, the existence of one cannot be completely ruled out without using significantly deeper exposure times or observing much more nearby SNe. The main limiting factor to future studies using this method is the scarcity of deep pre-explosion observations. We found that such \textit{Chandra }data exists only for $\sim6$ percent of SNe within $\sim$100 Mpc distance. Furthermore, our modelling indicates that HMXB-progenitor systems would likely be quite rare. However, these numbers could be increased if mass-transfer between the binary components is more efficient or supernova kicks are weaker than assumed. The long X-ray luminosity decline time of most SNe of these types \citep[see e.g.][]{ref15} limits the usefulness of post-explosion observations for recent SNe but could potentially allow the confirmation of candidate HMXB progenitors, once the SN has faded sufficiently.

\section*{Acknowledgements} \label{sect_acnowledgements}
We thank the referee for their helpful comments and feedback. This research has made use of data obtained from the Chandra Data Archive and the Chandra Source Catalog, and software provided by the Chandra X-ray Center (CXC) in the application packages CIAO, ChIPS, and Sherpa. This research has made use of the NASA/IPAC Extragalactic Database (NED) which is operated by the Jet Propulsion Laboratory, California Institute of Technology, under contract with the National Aeronautics and Space Administration. This research has made use of the VizieR catalogue access tool, CDS, Strasbourg, France. \citep{vizier} We also acknowledge the use of the Hyperleda database (http://leda.univ-lyon1.fr). The plots in this paper were made using matplotlib \citep{matplotlib}, http://matplotlib.org. This research has made use of NASA's Astrophysics Data System. This research was partly supported by the European Union FP7 programme through ERC grant number 320360.

\clearpage



\bibliographystyle{mnras} 
\bibliography{bibtex_uptd} 

\begin{thebibliography}{}
\makeatletter
\relax
\def\mn@urlcharsother{\let\do\@makeother \do\$\do\&\do\#\do\^\do\_\do\%\do\~}
\def\mn@doi{\begingroup\mn@urlcharsother \@ifnextchar [ {\mn@doi@}
  {\mn@doi@[]}}
\def\mn@doi@[#1]#2{\def\@tempa{#1}\ifx\@tempa\@empty \href
  {http://dx.doi.org/#2} {doi:#2}\else \href {http://dx.doi.org/#2} {#1}\fi
  \endgroup}
\def\mn@eprint#1#2{\mn@eprint@#1:#2::\@nil}
\def\mn@eprint@arXiv#1{\href {http://arxiv.org/abs/#1} {{\tt arXiv:#1}}}
\def\mn@eprint@dblp#1{\href {http://dblp.uni-trier.de/rec/bibtex/#1.xml}
  {dblp:#1}}
\def\mn@eprint@#1:#2:#3:#4\@nil{\def\@tempa {#1}\def\@tempb {#2}\def\@tempc
  {#3}\ifx \@tempc \@empty \let \@tempc \@tempb \let \@tempb \@tempa \fi \ifx
  \@tempb \@empty \def\@tempb {arXiv}\fi \@ifundefined
  {mn@eprint@\@tempb}{\@tempb:\@tempc}{\expandafter \expandafter \csname
  mn@eprint@\@tempb\endcsname \expandafter{\@tempc}}}

\bibitem[\protect\citeauthoryear{{Alard}}{{Alard}}{2000}]{alard_2000}
{Alard} C.,  2000, \mn@doi [\aaps] {10.1051/aas:2000214}, \href
  {http://adsabs.harvard.edu/abs/2000A%26AS..144..363A} {144, 363}

\bibitem[\protect\citeauthoryear{{Alard} \& {Lupton}}{{Alard} \&
  {Lupton}}{1998}]{alard_1998}
{Alard} C.,  {Lupton} R.~H.,  1998, \mn@doi [\apj] {10.1086/305984}, \href
  {http://adsabs.harvard.edu/abs/1998ApJ...503..325A} {503, 325}

\bibitem[\protect\citeauthoryear{{Aldering}, {Humphreys}  \&
  {Richmond}}{{Aldering} et~al.}{1994}]{ref20}
{Aldering} G.,  {Humphreys} R.~M.,   {Richmond} M.,  1994, \mn@doi [\aj]
  {10.1086/116886}, \href {http://adsabs.harvard.edu/abs/1994AJ....107..662A}
  {107, 662}

\bibitem[\protect\citeauthoryear{{Barbon}, {Buond{\'{\i}}}, {Cappellaro}  \&
  {Turatto}}{{Barbon} et~al.}{1999}]{asiago}
{Barbon} R.,  {Buond{\'{\i}}} V.,  {Cappellaro} E.,   {Turatto} M.,  1999,
  \mn@doi [\aaps] {10.1051/aas:1999404}, \href
  {http://adsabs.harvard.edu/abs/1999A%26AS..139..531B} {139, 531}

\bibitem[\protect\citeauthoryear{{Bauer} \& {Brandt}}{{Bauer} \&
  {Brandt}}{2004}]{ref13}
{Bauer} F.~E.,  {Brandt} W.~N.,  2004, \mn@doi [\apjl] {10.1086/380107}, \href
  {http://adsabs.harvard.edu/abs/2004ApJ...601L..67B} {601, L67}

\bibitem[\protect\citeauthoryear{{Bersten} et~al.,}{{Bersten}
  et~al.}{2014}]{bersten}
{Bersten} M.~C.,  et~al., 2014, \mn@doi [\aj] {10.1088/0004-6256/148/4/68},
  \href {http://adsabs.harvard.edu/abs/2014AJ....148...68B} {148, 68}

\bibitem[\protect\citeauthoryear{{Bietenholz} et~al.,}{{Bietenholz}
  et~al.}{2010}]{pos_2009bb}
{Bietenholz} M.~F.,  et~al., 2010, \mn@doi [\apj] {10.1088/0004-637X/725/1/4},
  \href {http://adsabs.harvard.edu/abs/2010ApJ...725....4B} {725, 4}

\bibitem[\protect\citeauthoryear{{Bufano} et~al.,}{{Bufano}
  et~al.}{2014}]{ref16}
{Bufano} F.,  et~al., 2014, \mn@doi [\mnras] {10.1093/mnras/stu065}, \href
  {http://adsabs.harvard.edu/abs/2014MNRAS.439.1807B} {439, 1807}

\bibitem[\protect\citeauthoryear{{Cao} et~al.,}{{Cao} et~al.}{2013}]{ref23}
{Cao} Y.,  et~al., 2013, \mn@doi [\apjl] {10.1088/2041-8205/775/1/L7}, \href
  {http://adsabs.harvard.edu/abs/2013ApJ...775L...7C} {775, L7}

\bibitem[\protect\citeauthoryear{{Carpano}, {Pollock}, {Wilms}, {Ehle}  \&
  {Schirmer}}{{Carpano} et~al.}{2007}]{ref12}
{Carpano} S.,  {Pollock} A.~M.~T.,  {Wilms} J.,  {Ehle} M.,   {Schirmer} M.,
  2007, \mn@doi [\aap] {10.1051/0004-6361:20066527}, \href
  {http://adsabs.harvard.edu/abs/2007A%26A...461L...9C} {461, L9}

\bibitem[\protect\citeauthoryear{{Chandra}, {Dwarkadas}, {Ray}, {Immler}  \&
  {Pooley}}{{Chandra} et~al.}{2009}]{ref30}
{Chandra} P.,  {Dwarkadas} V.~V.,  {Ray} A.,  {Immler} S.,   {Pooley} D.,
  2009, \mn@doi [\apj] {10.1088/0004-637X/699/1/388}, \href
  {http://adsabs.harvard.edu/abs/2009ApJ...699..388C} {699, 388}

\bibitem[\protect\citeauthoryear{{Crockett} et~al.,}{{Crockett}
  et~al.}{2008}]{crockett}
{Crockett} R.~M.,  et~al., 2008, \mn@doi [\mnras]
  {10.1111/j.1745-3933.2008.00540.x}, \href
  {http://adsabs.harvard.edu/abs/2008MNRAS.391L...5C} {391, L5}

\bibitem[\protect\citeauthoryear{{Crowther}}{{Crowther}}{2007}]{crowther}
{Crowther} P.~A.,  2007, \mn@doi [\araa]
  {10.1146/annurev.astro.45.051806.110615}, \href
  {http://adsabs.harvard.edu/abs/2007ARA%26A..45..177C} {45, 177}

\bibitem[\protect\citeauthoryear{{Dominici}, {Teixeira}, {Horvath}  \&
  {Holvorcem}}{{Dominici} et~al.}{1998}]{pos_1998bw}
{Dominici} T.~P.,  {Teixeira} R.,  {Horvath} J.~E.,   {Holvorcem} P.,  1998,
  \iaucirc, \href {http://adsabs.harvard.edu/abs/1998IAUC.6946....3D} {6946, 3}

\bibitem[\protect\citeauthoryear{{Dwarkadas} \& {Gruszko}}{{Dwarkadas} \&
  {Gruszko}}{2012}]{ref15}
{Dwarkadas} V.~V.,  {Gruszko} J.,  2012, \mn@doi [\mnras]
  {10.1111/j.1365-2966.2011.19808.x}, \href
  {http://adsabs.harvard.edu/abs/2012MNRAS.419.1515D} {419, 1515}

\bibitem[\protect\citeauthoryear{{Eldridge} \& {Stanway}}{{Eldridge} \&
  {Stanway}}{2009}]{bpass}
{Eldridge} J.~J.,  {Stanway} E.~R.,  2009, \mn@doi [\mnras]
  {10.1111/j.1365-2966.2009.15514.x}, \href
  {http://adsabs.harvard.edu/abs/2009MNRAS.400.1019E} {400, 1019}

\bibitem[\protect\citeauthoryear{{Eldridge}, {Izzard}  \& {Tout}}{{Eldridge}
  et~al.}{2008}]{new_bpass}
{Eldridge} J.~J.,  {Izzard} R.~G.,   {Tout} C.~A.,  2008, \mn@doi [\mnras]
  {10.1111/j.1365-2966.2007.12738.x}, \href
  {http://adsabs.harvard.edu/abs/2008MNRAS.384.1109E} {384, 1109}

\bibitem[\protect\citeauthoryear{{Eldridge}, {Fraser}, {Smartt}, {Maund}  \&
  {Crockett}}{{Eldridge} et~al.}{2013}]{ref4}
{Eldridge} J.~J.,  {Fraser} M.,  {Smartt} S.~J.,  {Maund} J.~R.,   {Crockett}
  R.~M.,  2013, \mn@doi [\mnras] {10.1093/mnras/stt1612}, \href
  {http://adsabs.harvard.edu/abs/2013MNRAS.436..774E} {436, 774}

\bibitem[\protect\citeauthoryear{{Eldridge}, {Fraser}, {Maund}  \&
  {Smartt}}{{Eldridge} et~al.}{2015}]{ref24}
{Eldridge} J.~J.,  {Fraser} M.,  {Maund} J.~R.,   {Smartt} S.~J.,  2015,
  \mn@doi [\mnras] {10.1093/mnras/stu2197}, \href
  {http://adsabs.harvard.edu/abs/2015MNRAS.446.2689E} {446, 2689}

\bibitem[\protect\citeauthoryear{{Ergon} et~al.,}{{Ergon} et~al.}{2014}]{dist2}
{Ergon} M.,  et~al., 2014, \mn@doi [\aap] {10.1051/0004-6361/201321850}, \href
  {http://adsabs.harvard.edu/abs/2014A%26A...562A..17E} {562, A17}

\bibitem[\protect\citeauthoryear{{Filippenko}}{{Filippenko}}{1997}]{ref22}
{Filippenko} A.~V.,  1997, \mn@doi [\araa] {10.1146/annurev.astro.35.1.309},
  \href {http://adsabs.harvard.edu/abs/1997ARA%26A..35..309F} {35, 309}

\bibitem[\protect\citeauthoryear{{Filippenko} \& {Chornock}}{{Filippenko} \&
  {Chornock}}{2001}]{ref40}
{Filippenko} A.~V.,  {Chornock} R.,  2001, \iaucirc, \href
  {http://adsabs.harvard.edu/abs/2001IAUC.7638....1F} {7638, 1}

\bibitem[\protect\citeauthoryear{{Fiorentino}, {Musella}  \&
  {Marconi}}{{Fiorentino} et~al.}{2013}]{dist14}
{Fiorentino} G.,  {Musella} I.,   {Marconi} M.,  2013, \mn@doi [\mnras]
  {10.1093/mnras/stt1193}, \href
  {http://adsabs.harvard.edu/abs/2013MNRAS.434.2866F} {434, 2866}

\bibitem[\protect\citeauthoryear{{Folatelli} et~al.,}{{Folatelli}
  et~al.}{2014}]{folatelli}
{Folatelli} G.,  et~al., 2014, \mn@doi [\apjl] {10.1088/2041-8205/793/2/L22},
  \href {http://adsabs.harvard.edu/abs/2014ApJ...793L..22F} {793, L22}

\bibitem[\protect\citeauthoryear{{Fremling} et~al.,}{{Fremling}
  et~al.}{2014}]{ref25}
{Fremling} C.,  et~al., 2014, \mn@doi [\aap] {10.1051/0004-6361/201423884},
  \href {http://adsabs.harvard.edu/abs/2014A%26A...565A.114F} {565, A114}

\bibitem[\protect\citeauthoryear{{Ganeshalingam}, {Li}  \&
  {Filippenko}}{{Ganeshalingam} et~al.}{2013}]{dist22}
{Ganeshalingam} M.,  {Li} W.,   {Filippenko} A.~V.,  2013, \mn@doi [\mnras]
  {10.1093/mnras/stt893}, \href
  {http://adsabs.harvard.edu/abs/2013MNRAS.433.2240G} {433, 2240}

\bibitem[\protect\citeauthoryear{{Gehrels}}{{Gehrels}}{1986}]{gehrels}
{Gehrels} N.,  1986, \mn@doi [\apj] {10.1086/164079}, \href
  {http://adsabs.harvard.edu/abs/1986ApJ...303..336G} {303, 336}

\bibitem[\protect\citeauthoryear{{Gerke}, {Kochanek}, {Prieto}, {Stanek}  \&
  {Macri}}{{Gerke} et~al.}{2011}]{dist1}
{Gerke} J.~R.,  {Kochanek} C.~S.,  {Prieto} J.~L.,  {Stanek} K.~Z.,   {Macri}
  L.~M.,  2011, \mn@doi [\apj] {10.1088/0004-637X/743/2/176}, \href
  {http://adsabs.harvard.edu/abs/2011ApJ...743..176G} {743, 176}

\bibitem[\protect\citeauthoryear{{Gerke}, {Kochanek}  \& {Stanek}}{{Gerke}
  et~al.}{2015}]{gerke_dsn}
{Gerke} J.~R.,  {Kochanek} C.~S.,   {Stanek} K.~Z.,  2015, \mn@doi [\mnras]
  {10.1093/mnras/stv776}, \href
  {http://adsabs.harvard.edu/abs/2015MNRAS.450.3289G} {450, 3289}

\bibitem[\protect\citeauthoryear{{Goodall}, {Alouani-Bibi}  \&
  {Blundell}}{{Goodall} et~al.}{2011}]{SNR_goodall}
{Goodall} P.~T.,  {Alouani-Bibi} F.,   {Blundell} K.~M.,  2011, \mn@doi
  [\mnras] {10.1111/j.1365-2966.2011.18388.x}, \href
  {http://adsabs.harvard.edu/abs/2011MNRAS.414.2838G} {414, 2838}

\bibitem[\protect\citeauthoryear{{Hamaguchi} et~al.,}{{Hamaguchi}
  et~al.}{2014a}]{etacar}
{Hamaguchi} K.,  et~al., 2014a, \mn@doi [\apj] {10.1088/0004-637X/784/2/125},
  \href {http://adsabs.harvard.edu/abs/2014ApJ...784..125H} {784, 125}

\bibitem[\protect\citeauthoryear{{Hamaguchi} et~al.,}{{Hamaguchi}
  et~al.}{2014b}]{hamaguchi}
{Hamaguchi} K.,  et~al., 2014b, \mn@doi [\apj] {10.1088/0004-637X/795/2/119},
  \href {http://adsabs.harvard.edu/abs/2014ApJ...795..119H} {795, 119}

\bibitem[\protect\citeauthoryear{{Heinz} et~al.,}{{Heinz}
  et~al.}{2013}]{SNR_heinz}
{Heinz} S.,  et~al., 2013, \mn@doi [\apj] {10.1088/0004-637X/779/2/171}, \href
  {http://adsabs.harvard.edu/abs/2013ApJ...779..171H} {779, 171}

\bibitem[\protect\citeauthoryear{{H{\'e}nault-Brunet}
  et~al.,}{{H{\'e}nault-Brunet} et~al.}{2012}]{SNR_brunet}
{H{\'e}nault-Brunet} V.,  et~al., 2012, \mn@doi [\mnras]
  {10.1111/j.1745-3933.2011.01183.x}, \href
  {http://adsabs.harvard.edu/abs/2012MNRAS.420L..13H} {420, L13}

\bibitem[\protect\citeauthoryear{{Hicken}, {Wood-Vasey}, {Blondin}, {Challis},
  {Jha}, {Kelly}, {Rest}  \& {Kirshner}}{{Hicken} et~al.}{2009}]{dist24}
{Hicken} M.,  {Wood-Vasey} W.~M.,  {Blondin} S.,  {Challis} P.,  {Jha} S.,
  {Kelly} P.~L.,  {Rest} A.,   {Kirshner} R.~P.,  2009, \mn@doi [\apj]
  {10.1088/0004-637X/700/2/1097}, \href
  {http://adsabs.harvard.edu/abs/2009ApJ...700.1097H} {700, 1097}

\bibitem[\protect\citeauthoryear{{Hunter}}{{Hunter}}{2007}]{matplotlib}
{Hunter} J.~D.,  2007, \mn@doi [Computing in Science and Engineering]
  {10.1109/MCSE.2007.55}, \href
  {http://adsabs.harvard.edu/abs/2007CSE.....9...90H} {9, 90}

\bibitem[\protect\citeauthoryear{{Hunter} et~al.,}{{Hunter}
  et~al.}{2009}]{dist3}
{Hunter} D.~J.,  et~al., 2009, \mn@doi [\aap] {10.1051/0004-6361/200912896},
  \href {http://adsabs.harvard.edu/abs/2009A%26A...508..371H} {508, 371}

\bibitem[\protect\citeauthoryear{{Hurley}, {Tout}  \& {Pols}}{{Hurley}
  et~al.}{2002}]{hurley}
{Hurley} J.~R.,  {Tout} C.~A.,   {Pols} O.~R.,  2002, \mn@doi [\mnras]
  {10.1046/j.1365-8711.2002.05038.x}, \href
  {http://adsabs.harvard.edu/abs/2002MNRAS.329..897H} {329, 897}

\bibitem[\protect\citeauthoryear{{Immler}, {Wilson}  \& {Terashima}}{{Immler}
  et~al.}{2002}]{ref36}
{Immler} S.,  {Wilson} A.~S.,   {Terashima} Y.,  2002, \mn@doi [\apjl]
  {10.1086/341935}, \href {http://adsabs.harvard.edu/abs/2002ApJ...573L..27I}
  {573, L27}

\bibitem[\protect\citeauthoryear{{Jha}, {Riess}  \& {Kirshner}}{{Jha}
  et~al.}{2007}]{dist26}
{Jha} S.,  {Riess} A.~G.,   {Kirshner} R.~P.,  2007, \mn@doi [\apj]
  {10.1086/512054}, \href {http://adsabs.harvard.edu/abs/2007ApJ...659..122J}
  {659, 122}

\bibitem[\protect\citeauthoryear{{Kangas}, {Mattila}, {Kankare}, {Kotilainen},
  {V{\"a}is{\"a}nen}, {Greimel}  \& {Takalo}}{{Kangas}
  et~al.}{2013}]{kangas_2013}
{Kangas} T.,  {Mattila} S.,  {Kankare} E.,  {Kotilainen} J.~K.,
  {V{\"a}is{\"a}nen} P.,  {Greimel} R.,   {Takalo} A.,  2013, \mn@doi [\mnras]
  {10.1093/mnras/stt1833}, \href
  {http://adsabs.harvard.edu/abs/2013MNRAS.436.3464K} {436, 3464}

\bibitem[\protect\citeauthoryear{{Karachentsev}, {Makarov}  \&
  {Kaisina}}{{Karachentsev} et~al.}{2013}]{dist5}
{Karachentsev} I.~D.,  {Makarov} D.~I.,   {Kaisina} E.~I.,  2013, \mn@doi [\aj]
  {10.1088/0004-6256/145/4/101}, \href
  {http://adsabs.harvard.edu/abs/2013AJ....145..101K} {145, 101}

\bibitem[\protect\citeauthoryear{{Kouveliotou} et~al.,}{{Kouveliotou}
  et~al.}{2004}]{ref28}
{Kouveliotou} C.,  et~al., 2004, \mn@doi [\apj] {10.1086/420878}, \href
  {http://adsabs.harvard.edu/abs/2004ApJ...608..872K} {608, 872}

\bibitem[\protect\citeauthoryear{{Kuncarayakti} et~al.,}{{Kuncarayakti}
  et~al.}{2015}]{nebular_phase}
{Kuncarayakti} H.,  et~al., 2015, \mn@doi [\aap] {10.1051/0004-6361/201425604},
  \href {http://adsabs.harvard.edu/abs/2015A%26A...579A..95K} {579, A95}

\bibitem[\protect\citeauthoryear{{Lennarz}, {Altmann}  \& {Wiebusch}}{{Lennarz}
  et~al.}{2012}]{lennarz}
{Lennarz} D.,  {Altmann} D.,   {Wiebusch} C.,  2012, \mn@doi [\aap]
  {10.1051/0004-6361/201117666}, \href
  {http://adsabs.harvard.edu/abs/2012A%26A...538A.120L} {538, A120}

\bibitem[\protect\citeauthoryear{{Lyman}, {Bersier}, {James}, {Mazzali},
  {Eldridge}, {Fraser}  \& {Pian}}{{Lyman} et~al.}{2014}]{lyman}
{Lyman} J.,  {Bersier} D.,  {James} P.,  {Mazzali} P.,  {Eldridge} J.,
  {Fraser} M.,   {Pian} E.,  2014, preprint, \href
  {http://adsabs.harvard.edu/abs/2014arXiv1406.3667L} {} (\mn@eprint {arXiv}
  {1406.3667})

\bibitem[\protect\citeauthoryear{{Maccarone}, {Lehmer}, {Leyder}, {Antoniou},
  {Hornschemeier}, {Ptak}, {Wik}  \& {Zezas}}{{Maccarone} et~al.}{2014}]{ref11}
{Maccarone} T.~J.,  {Lehmer} B.~D.,  {Leyder} J.~C.,  {Antoniou} V.,
  {Hornschemeier} A.,  {Ptak} A.,  {Wik} D.,   {Zezas} A.,  2014, \mn@doi
  [\mnras] {10.1093/mnras/stu167}, \href
  {http://adsabs.harvard.edu/abs/2014MNRAS.439.3064M} {439, 3064}

\bibitem[\protect\citeauthoryear{{Maeda}, {Katsuda}, {Bamba}, {Terada}  \&
  {Fukazawa}}{{Maeda} et~al.}{2014}]{ref32}
{Maeda} K.,  {Katsuda} S.,  {Bamba} A.,  {Terada} Y.,   {Fukazawa} Y.,  2014,
  \mn@doi [\apj] {10.1088/0004-637X/785/2/95}, \href
  {http://adsabs.harvard.edu/abs/2014ApJ...785...95M} {785, 95}

\bibitem[\protect\citeauthoryear{{Mandel}, {Wood-Vasey}, {Friedman}  \&
  {Kirshner}}{{Mandel} et~al.}{2009}]{dist19}
{Mandel} K.~S.,  {Wood-Vasey} W.~M.,  {Friedman} A.~S.,   {Kirshner} R.~P.,
  2009, \mn@doi [\apj] {10.1088/0004-637X/704/1/629}, \href
  {http://adsabs.harvard.edu/abs/2009ApJ...704..629M} {704, 629}

\bibitem[\protect\citeauthoryear{{Mandel}, {Narayan}  \& {Kirshner}}{{Mandel}
  et~al.}{2011}]{dist18}
{Mandel} K.~S.,  {Narayan} G.,   {Kirshner} R.~P.,  2011, \mn@doi [\apj]
  {10.1088/0004-637X/731/2/120}, \href
  {http://adsabs.harvard.edu/abs/2011ApJ...731..120M} {731, 120}

\bibitem[\protect\citeauthoryear{{Marcaide} et~al.,}{{Marcaide}
  et~al.}{1993}]{pos_1993J}
{Marcaide} J.~M.,  et~al., 1993, \iaucirc, \href
  {http://adsabs.harvard.edu/abs/1993IAUC.5820....2M} {5820, 2}

\bibitem[\protect\citeauthoryear{{Margutti}, {Soderberg}, {Chakraborti},
  {Drout}, {Kamble}, {Milisavljevic}, {Sanders}  \& {Zauderer}}{{Margutti}
  et~al.}{2013}]{ref33}
{Margutti} R.,  {Soderberg} A.,  {Chakraborti} S.,  {Drout} M.,  {Kamble} A.,
  {Milisavljevic} D.,  {Sanders} N.,   {Zauderer} A.,  2013, The Astronomer's
  Telegram, \href {http://adsabs.harvard.edu/abs/2013ATel.4944....1M} {4944, 1}

\bibitem[\protect\citeauthoryear{{Mattila}, {Meikle}  \& {Greimel}}{{Mattila}
  et~al.}{2004}]{mattila_starburst}
{Mattila} S.,  {Meikle} W.~P.~S.,   {Greimel} R.,  2004, \mn@doi [\nar]
  {10.1016/j.newar.2003.12.033}, \href
  {http://adsabs.harvard.edu/abs/2004NewAR..48..595M} {48, 595}

\bibitem[\protect\citeauthoryear{{Mattila}, {Monard}  \& {Li}}{{Mattila}
  et~al.}{2005}]{2005U_pos_Li}
{Mattila} S.,  {Monard} L.~A.~G.,   {Li} W.,  2005, \iaucirc, \href
  {http://adsabs.harvard.edu/abs/2005IAUC.8473....2M} {8473, 2}

\bibitem[\protect\citeauthoryear{{Mattila} et~al.,}{{Mattila}
  et~al.}{2012}]{dist11}
{Mattila} S.,  et~al., 2012, \mn@doi [\apj] {10.1088/0004-637X/756/2/111},
  \href {http://adsabs.harvard.edu/abs/2012ApJ...756..111M} {756, 111}

\bibitem[\protect\citeauthoryear{{Maund} \& {Smartt}}{{Maund} \&
  {Smartt}}{2009}]{maund_smartt}
{Maund} J.~R.,  {Smartt} S.~J.,  2009, \mn@doi [Science]
  {10.1126/science.1170198}, \href
  {http://adsabs.harvard.edu/abs/2009Sci...324..486M} {324, 486}

\bibitem[\protect\citeauthoryear{{Maund}, {Smartt}  \& {Schweizer}}{{Maund}
  et~al.}{2005}]{maund_2005}
{Maund} J.~R.,  {Smartt} S.~J.,   {Schweizer} F.,  2005, \mn@doi [\apjl]
  {10.1086/491620}, \href {http://adsabs.harvard.edu/abs/2005ApJ...630L..33M}
  {630, L33}

\bibitem[\protect\citeauthoryear{{Maund} et~al.,}{{Maund} et~al.}{2011}]{ref21}
{Maund} J.~R.,  et~al., 2011, \mn@doi [\apjl] {10.1088/2041-8205/739/2/L37},
  \href {http://adsabs.harvard.edu/abs/2011ApJ...739L..37M} {739, L37}

\bibitem[\protect\citeauthoryear{{Maund} et~al.,}{{Maund}
  et~al.}{2015}]{maund_2015}
{Maund} J.~R.,  et~al., 2015, \mn@doi [\mnras] {10.1093/mnras/stv2098}, \href
  {http://adsabs.harvard.edu/abs/2015MNRAS.454.2580M} {454, 2580}

\bibitem[\protect\citeauthoryear{{McCollough}, {Smith}  \&
  {Valencic}}{{McCollough} et~al.}{2013}]{ref14}
{McCollough} M.~L.,  {Smith} R.~K.,   {Valencic} L.~A.,  2013, \mn@doi [\apj]
  {10.1088/0004-637X/762/1/2}, \href
  {http://adsabs.harvard.edu/abs/2013ApJ...762....2M} {762, 2}

\bibitem[\protect\citeauthoryear{{Mineo}, {Gilfanov}  \& {Sunyaev}}{{Mineo}
  et~al.}{2012}]{mineo}
{Mineo} S.,  {Gilfanov} M.,   {Sunyaev} R.,  2012, \mn@doi [\mnras]
  {10.1111/j.1365-2966.2011.19862.x}, \href
  {http://adsabs.harvard.edu/abs/2012MNRAS.419.2095M} {419, 2095}

\bibitem[\protect\citeauthoryear{{Mitsuda} et~al.,}{{Mitsuda}
  et~al.}{1984}]{ref27}
{Mitsuda} K.,  et~al., 1984, \pasj, \href
  {http://adsabs.harvard.edu/abs/1984PASJ...36..741M} {36, 741}

\bibitem[\protect\citeauthoryear{{Modjaz} et~al.,}{{Modjaz}
  et~al.}{2009}]{ref7}
{Modjaz} M.,  et~al., 2009, \mn@doi [\apj] {10.1088/0004-637X/702/1/226}, \href
  {http://adsabs.harvard.edu/abs/2009ApJ...702..226M} {702, 226}

\bibitem[\protect\citeauthoryear{{Monet}}{{Monet}}{1998}]{USNO}
{Monet} D.,  1998, {USNO-A2.0}

\bibitem[\protect\citeauthoryear{{Nasonova}, {de Freitas Pacheco}  \&
  {Karachentsev}}{{Nasonova} et~al.}{2011}]{dist8}
{Nasonova} O.~G.,  {de Freitas Pacheco} J.~A.,   {Karachentsev} I.~D.,  2011,
  \mn@doi [\aap] {10.1051/0004-6361/201016004}, \href
  {http://adsabs.harvard.edu/abs/2011A%26A...532A.104N} {532, A104}

\bibitem[\protect\citeauthoryear{{Naz{\'e}}, {Rauw}  \&
  {Hutsem{\'e}kers}}{{Naz{\'e}} et~al.}{2012}]{lbv_sur}
{Naz{\'e}} Y.,  {Rauw} G.,   {Hutsem{\'e}kers} D.,  2012, \mn@doi [\aap]
  {10.1051/0004-6361/201118040}, \href
  {http://adsabs.harvard.edu/abs/2012A%26A...538A..47N} {538, A47}

\bibitem[\protect\citeauthoryear{{Nelemans}, {Voss}, {Nielsen}  \&
  {Roelofs}}{{Nelemans} et~al.}{2010}]{ref2}
{Nelemans} G.,  {Voss} R.,  {Nielsen} M.~T.~B.,   {Roelofs} G.,  2010, \mn@doi
  [\mnras] {10.1111/j.1745-3933.2010.00861.x}, \href
  {http://adsabs.harvard.edu/abs/2010MNRAS.405L..71N} {405, L71}

\bibitem[\protect\citeauthoryear{{Nielsen}, {Voss}  \& {Nelemans}}{{Nielsen}
  et~al.}{2012}]{ref1}
{Nielsen} M.~T.~B.,  {Voss} R.,   {Nelemans} G.,  2012, \mn@doi [\mnras]
  {10.1111/j.1365-2966.2012.21596.x}, \href
  {http://adsabs.harvard.edu/abs/2012MNRAS.426.2668N} {426, 2668}

\bibitem[\protect\citeauthoryear{{Nielsen}, {Voss}  \& {Nelemans}}{{Nielsen}
  et~al.}{2013}]{ref37}
{Nielsen} M.~T.~B.,  {Voss} R.,   {Nelemans} G.,  2013, \mn@doi [\mnras]
  {10.1093/mnras/stt1250}, \href
  {http://adsabs.harvard.edu/abs/2013MNRAS.435..187N} {435, 187}

\bibitem[\protect\citeauthoryear{{Ochsenbein}, {Bauer}  \&
  {Marcout}}{{Ochsenbein} et~al.}{2000}]{vizier}
{Ochsenbein} F.,  {Bauer} P.,   {Marcout} J.,  2000, \mn@doi [\aaps]
  {10.1051/aas:2000169}, \href
  {http://adsabs.harvard.edu/abs/2000A%26AS..143...23O} {143, 23}

\bibitem[\protect\citeauthoryear{{Pastorello} et~al.,}{{Pastorello}
  et~al.}{2008}]{dist6}
{Pastorello} A.,  et~al., 2008, \mn@doi [\mnras]
  {10.1111/j.1365-2966.2008.13618.x}, \href
  {http://adsabs.harvard.edu/abs/2008MNRAS.389..955P} {389, 955}

\bibitem[\protect\citeauthoryear{{Paturel}, {Petit}, {Prugniel}, {Theureau},
  {Rousseau}, {Brouty}, {Dubois}  \& {Cambr{\'e}sy}}{{Paturel}
  et~al.}{2003}]{refleda}
{Paturel} G.,  {Petit} C.,  {Prugniel} P.,  {Theureau} G.,  {Rousseau} J.,
  {Brouty} M.,  {Dubois} P.,   {Cambr{\'e}sy} L.,  2003, \mn@doi [\aap]
  {10.1051/0004-6361:20031411}, \href
  {http://adsabs.harvard.edu/abs/2003A%26A...412...45P} {412, 45}

\bibitem[\protect\citeauthoryear{{Perna}, {Soria}, {Pooley}  \&
  {Stella}}{{Perna} et~al.}{2008}]{ref34}
{Perna} R.,  {Soria} R.,  {Pooley} D.,   {Stella} L.,  2008, \mn@doi [\mnras]
  {10.1111/j.1365-2966.2007.12821.x}, \href
  {http://adsabs.harvard.edu/abs/2008MNRAS.384.1638P} {384, 1638}

\bibitem[\protect\citeauthoryear{{Podsiadlowski}, {Joss}  \&
  {Hsu}}{{Podsiadlowski} et~al.}{1992}]{ref5}
{Podsiadlowski} P.,  {Joss} P.~C.,   {Hsu} J.~J.~L.,  1992, \mn@doi [\apj]
  {10.1086/171341}, \href {http://adsabs.harvard.edu/abs/1992ApJ...391..246P}
  {391, 246}

\bibitem[\protect\citeauthoryear{{Pooley} \& {Lewin}}{{Pooley} \&
  {Lewin}}{2003}]{2003bg_disc}
{Pooley} D.,  {Lewin} W.~H.~G.,  2003, \iaucirc, \href
  {http://adsabs.harvard.edu/abs/2003IAUC.8110....2P} {8110, 2}

\bibitem[\protect\citeauthoryear{{Pounds} \& {King}}{{Pounds} \&
  {King}}{2013}]{dist9}
{Pounds} K.~A.,  {King} A.~R.,  2013, \mn@doi [\mnras] {10.1093/mnras/stt807},
  \href {http://adsabs.harvard.edu/abs/2013MNRAS.433.1369P} {433, 1369}

\bibitem[\protect\citeauthoryear{{Poutanen}, {Fabrika}, {Valeev}, {Sholukhova}
  \& {Greiner}}{{Poutanen} et~al.}{2013}]{juri_ulx}
{Poutanen} J.,  {Fabrika} S.,  {Valeev} A.~F.,  {Sholukhova} O.,   {Greiner}
  J.,  2013, \mn@doi [\mnras] {10.1093/mnras/stt487}, \href
  {http://adsabs.harvard.edu/abs/2013MNRAS.432..506P} {432, 506}

\bibitem[\protect\citeauthoryear{{Poznanski} et~al.,}{{Poznanski}
  et~al.}{2009}]{dist15}
{Poznanski} D.,  et~al., 2009, \mn@doi [\apj] {10.1088/0004-637X/694/2/1067},
  \href {http://adsabs.harvard.edu/abs/2009ApJ...694.1067P} {694, 1067}

\bibitem[\protect\citeauthoryear{{Predehl} \& {Schmitt}}{{Predehl} \&
  {Schmitt}}{1995}]{predehl}
{Predehl} P.,  {Schmitt} J.~H.~M.~M.,  1995, \aap, \href
  {http://adsabs.harvard.edu/abs/1995A%26A...293..889P} {293, 889}

\bibitem[\protect\citeauthoryear{{Prieto}, {Rest}  \& {Suntzeff}}{{Prieto}
  et~al.}{2006}]{dist25}
{Prieto} J.~L.,  {Rest} A.,   {Suntzeff} N.~B.,  2006, \mn@doi [\apj]
  {10.1086/504307}, \href {http://adsabs.harvard.edu/abs/2006ApJ...647..501P}
  {647, 501}

\bibitem[\protect\citeauthoryear{{Reindl}, {Tammann}, {Sandage}  \&
  {Saha}}{{Reindl} et~al.}{2005}]{dist27}
{Reindl} B.,  {Tammann} G.~A.,  {Sandage} A.,   {Saha} A.,  2005, \mn@doi
  [\apj] {10.1086/429218}, \href
  {http://adsabs.harvard.edu/abs/2005ApJ...624..532R} {624, 532}

\bibitem[\protect\citeauthoryear{{Reynolds}, {Fraser}  \& {Gilmore}}{{Reynolds}
  et~al.}{2015}]{reynolds_dsn}
{Reynolds} T.~M.,  {Fraser} M.,   {Gilmore} G.,  2015, \mn@doi [\mnras]
  {10.1093/mnras/stv1809}, \href
  {http://adsabs.harvard.edu/abs/2015MNRAS.453.2885R} {453, 2885}

\bibitem[\protect\citeauthoryear{{Roming} et~al.,}{{Roming}
  et~al.}{2009}]{roming}
{Roming} P.~W.~A.,  et~al., 2009, \mn@doi [\apjl]
  {10.1088/0004-637X/704/2/L118}, \href
  {http://adsabs.harvard.edu/abs/2009ApJ...704L.118R} {704, L118}

\bibitem[\protect\citeauthoryear{{Rupen}, {Sramek}, {van Dyk}, {Weiler},
  {Panagia}, {Richmond}, {Filippenko}  \& {Treffers}}{{Rupen}
  et~al.}{1994}]{pos_1994I}
{Rupen} M.~P.,  {Sramek} R.~A.,  {van Dyk} S.~D.,  {Weiler} K.~W.,  {Panagia}
  N.,  {Richmond} M.~W.,  {Filippenko} A.~V.,   {Treffers} R.~R.,  1994,
  \iaucirc, \href {http://adsabs.harvard.edu/abs/1994IAUC.5963....1R} {5963, 1}

\bibitem[\protect\citeauthoryear{{Russell}}{{Russell}}{2002}]{dist10}
{Russell} D.~G.,  2002, \mn@doi [\apj] {10.1086/337917}, \href
  {http://adsabs.harvard.edu/abs/2002ApJ...565..681R} {565, 681}

\bibitem[\protect\citeauthoryear{{Ryder}, {Sadler}, {Subrahmanyan}, {Weiler},
  {Panagia}  \& {Stockdale}}{{Ryder} et~al.}{2004}]{ref17}
{Ryder} S.~D.,  {Sadler} E.~M.,  {Subrahmanyan} R.,  {Weiler} K.~W.,  {Panagia}
  N.,   {Stockdale} C.,  2004, \mn@doi [\mnras]
  {10.1111/j.1365-2966.2004.07589.x}, \href
  {http://adsabs.harvard.edu/abs/2004MNRAS.349.1093R} {349, 1093}

\bibitem[\protect\citeauthoryear{{Ryder}, {Murrowood}  \& {Stathakis}}{{Ryder}
  et~al.}{2006}]{ref19}
{Ryder} S.~D.,  {Murrowood} C.~E.,   {Stathakis} R.~A.,  2006, \mn@doi [\mnras]
  {10.1111/j.1745-3933.2006.00168.x}, \href
  {http://adsabs.harvard.edu/abs/2006MNRAS.369L..32R} {369, L32}

\bibitem[\protect\citeauthoryear{{Saha}, {Thim}, {Tammann}, {Reindl}  \&
  {Sandage}}{{Saha} et~al.}{2006}]{dist4}
{Saha} A.,  {Thim} F.,  {Tammann} G.~A.,  {Reindl} B.,   {Sandage} A.,  2006,
  \mn@doi [\apjs] {10.1086/503800}, \href
  {http://adsabs.harvard.edu/abs/2006ApJS..165..108S} {165, 108}

\bibitem[\protect\citeauthoryear{{Sana} et~al.,}{{Sana} et~al.}{2012}]{sana}
{Sana} H.,  et~al., 2012, \mn@doi [Science] {10.1126/science.1223344}, \href
  {http://adsabs.harvard.edu/abs/2012Sci...337..444S} {337, 444}

\bibitem[\protect\citeauthoryear{{Schlegel} \& {Ryder}}{{Schlegel} \&
  {Ryder}}{2002}]{ref31}
{Schlegel} E.~M.,  {Ryder} S.,  2002, \iaucirc, \href
  {http://adsabs.harvard.edu/abs/2002IAUC.7913....1S} {7913, 1}

\bibitem[\protect\citeauthoryear{{Seward}, {Charles}, {Foster}, {Dickel},
  {Romero}, {Edwards}, {Perry}  \& {Williams}}{{Seward}
  et~al.}{2012}]{SNR_seward}
{Seward} F.~D.,  {Charles} P.~A.,  {Foster} D.~L.,  {Dickel} J.~R.,  {Romero}
  P.~S.,  {Edwards} Z.~I.,  {Perry} M.,   {Williams} R.~M.,  2012, \mn@doi
  [\apj] {10.1088/0004-637X/759/2/123}, \href
  {http://adsabs.harvard.edu/abs/2012ApJ...759..123S} {759, 123}

\bibitem[\protect\citeauthoryear{{Smartt}}{{Smartt}}{2009}]{ref3}
{Smartt} S.~J.,  2009, \mn@doi [\araa] {10.1146/annurev-astro-082708-101737},
  \href {http://adsabs.harvard.edu/abs/2009ARA%26A..47...63S} {47, 63}

\bibitem[\protect\citeauthoryear{{Smartt}}{{Smartt}}{2015}]{newsmartt}
{Smartt} S.~J.,  2015, \mn@doi [\pasa] {10.1017/pasa.2015.17}, \href
  {http://adsabs.harvard.edu/abs/2015PASA...32...16S} {32, 16}

\bibitem[\protect\citeauthoryear{{Smith}, {Li}, {Filippenko}  \&
  {Chornock}}{{Smith} et~al.}{2011}]{ref9}
{Smith} N.,  {Li} W.,  {Filippenko} A.~V.,   {Chornock} R.,  2011, \mn@doi
  [\mnras] {10.1111/j.1365-2966.2011.17229.x}, \href
  {http://adsabs.harvard.edu/abs/2011MNRAS.412.1522S} {412, 1522}

\bibitem[\protect\citeauthoryear{{Soderberg}, {Kulkarni}, {Berger},
  {Chevalier}, {Frail}, {Fox}  \& {Walker}}{{Soderberg}
  et~al.}{2005}]{dist28_b}
{Soderberg} A.~M.,  {Kulkarni} S.~R.,  {Berger} E.,  {Chevalier} R.~A.,
  {Frail} D.~A.,  {Fox} D.~B.,   {Walker} R.~C.,  2005, \mn@doi [\apj]
  {10.1086/427649}, \href {http://adsabs.harvard.edu/abs/2005ApJ...621..908S}
  {621, 908}

\bibitem[\protect\citeauthoryear{{Soderberg} et~al.,}{{Soderberg}
  et~al.}{2008}]{ref6}
{Soderberg} A.~M.,  et~al., 2008, \mn@doi [\nat] {10.1038/nature06997}, \href
  {http://adsabs.harvard.edu/abs/2008Natur.453..469S} {453, 469}

\bibitem[\protect\citeauthoryear{{Springob}, {Masters}, {Haynes}, {Giovanelli}
  \& {Marinoni}}{{Springob} et~al.}{2009}]{dist16}
{Springob} C.~M.,  {Masters} K.~L.,  {Haynes} M.~P.,  {Giovanelli} R.,
  {Marinoni} C.,  2009, \mn@doi [\apjs] {10.1088/0067-0049/182/1/474}, \href
  {http://adsabs.harvard.edu/abs/2009ApJS..182..474S} {182, 474}

\bibitem[\protect\citeauthoryear{{Svirski} \& {Nakar}}{{Svirski} \&
  {Nakar}}{2014}]{ref8}
{Svirski} G.,  {Nakar} E.,  2014, \mn@doi [\apjl]
  {10.1088/2041-8205/788/1/L14}, \href
  {http://adsabs.harvard.edu/abs/2014ApJ...788L..14S} {788, L14}

\bibitem[\protect\citeauthoryear{{Takanashi}, {Doi}  \& {Yasuda}}{{Takanashi}
  et~al.}{2008}]{dist21}
{Takanashi} N.,  {Doi} M.,   {Yasuda} N.,  2008, \mn@doi [\mnras]
  {10.1111/j.1365-2966.2008.13694.x}, \href
  {http://adsabs.harvard.edu/abs/2008MNRAS.389.1577T} {389, 1577}

\bibitem[\protect\citeauthoryear{{Terry}, {Paturel}  \& {Ekholm}}{{Terry}
  et~al.}{2002}]{dist12}
{Terry} J.~N.,  {Paturel} G.,   {Ekholm} T.,  2002, \mn@doi [\aap]
  {10.1051/0004-6361:20021018}, \href
  {http://adsabs.harvard.edu/abs/2002A%26A...393...57T} {393, 57}

\bibitem[\protect\citeauthoryear{{Theureau}, {Hanski}, {Coudreau}, {Hallet}  \&
  {Martin}}{{Theureau} et~al.}{2007}]{dist13}
{Theureau} G.,  {Hanski} M.~O.,  {Coudreau} N.,  {Hallet} N.,   {Martin} J.-M.,
   2007, \mn@doi [\aap] {10.1051/0004-6361:20066187}, \href
  {http://adsabs.harvard.edu/abs/2007A%26A...465...71T} {465, 71}

\bibitem[\protect\citeauthoryear{{Tsvetkov}}{{Tsvetkov}}{1985}]{1983I_disc}
{Tsvetkov} D.~Y.,  1985, \sovast, \href
  {http://adsabs.harvard.edu/abs/1985SvA....29..211T} {29, 211}

\bibitem[\protect\citeauthoryear{{Tsvetkov}, {Pavlyuk}  \&
  {Bartunov}}{{Tsvetkov} et~al.}{2004}]{sternberg}
{Tsvetkov} D.~Y.,  {Pavlyuk} N.~N.,   {Bartunov} O.~S.,  2004, PAZh, 30, 803

\bibitem[\protect\citeauthoryear{{Tully}, {Rizzi}, {Shaya}, {Courtois},
  {Makarov}  \& {Jacobs}}{{Tully} et~al.}{2009}]{dist7}
{Tully} R.~B.,  {Rizzi} L.,  {Shaya} E.~J.,  {Courtois} H.~M.,  {Makarov}
  D.~I.,   {Jacobs} B.~A.,  2009, \mn@doi [\aj] {10.1088/0004-6256/138/2/323},
  \href {http://adsabs.harvard.edu/abs/2009AJ....138..323T} {138, 323}

\bibitem[\protect\citeauthoryear{{Valenti} et~al.,}{{Valenti}
  et~al.}{2011}]{ref10}
{Valenti} S.,  et~al., 2011, \mn@doi [\mnras]
  {10.1111/j.1365-2966.2011.19262.x}, \href
  {http://adsabs.harvard.edu/abs/2011MNRAS.416.3138V} {416, 3138}

\bibitem[\protect\citeauthoryear{{Van Dyk}, {Li}  \& {Filippenko}}{{Van Dyk}
  et~al.}{2003}]{vandykpos}
{Van Dyk} S.~D.,  {Li} W.,   {Filippenko} A.~V.,  2003, \mn@doi [\pasp]
  {10.1086/345748}, \href {http://adsabs.harvard.edu/abs/2003PASP..115....1V}
  {115, 1}

\bibitem[\protect\citeauthoryear{{Van Dyk} et~al.,}{{Van Dyk}
  et~al.}{2011}]{vandyk_2011dh}
{Van Dyk} S.~D.,  et~al., 2011, \mn@doi [\apjl] {10.1088/2041-8205/741/2/L28},
  \href {http://adsabs.harvard.edu/abs/2011ApJ...741L..28V} {741, L28}

\bibitem[\protect\citeauthoryear{{Vanbeveren}, {De Donder}, {Van Bever}, {Van
  Rensbergen}  \& {De Loore}}{{Vanbeveren} et~al.}{1998}]{vanbeveren_1998}
{Vanbeveren} D.,  {De Donder} E.,  {Van Bever} J.,  {Van Rensbergen} W.,   {De
  Loore} C.,  1998, \mn@doi [\na] {10.1016/S1384-1076(98)00020-7}, \href
  {http://adsabs.harvard.edu/abs/1998NewA....3..443V} {3, 443}

\bibitem[\protect\citeauthoryear{{Voss}, {Nielsen}, {Nelemans}, {Fraser}  \&
  {Smartt}}{{Voss} et~al.}{2011}]{ref26}
{Voss} R.,  {Nielsen} M.~T.~B.,  {Nelemans} G.,  {Fraser} M.,   {Smartt} S.~J.,
   2011, \mn@doi [\mnras] {10.1111/j.1745-3933.2011.01157.x}, \href
  {http://adsabs.harvard.edu/abs/2011MNRAS.418L.124V} {418, L124}

\bibitem[\protect\citeauthoryear{{Wang}, {Wang}, {Pain}, {Zhou}  \&
  {Li}}{{Wang} et~al.}{2006}]{dist20}
{Wang} X.,  {Wang} L.,  {Pain} R.,  {Zhou} X.,   {Li} Z.,  2006, \mn@doi [\apj]
  {10.1086/504312}, \href {http://adsabs.harvard.edu/abs/2006ApJ...645..488W}
  {645, 488}

\bibitem[\protect\citeauthoryear{{Weyant}, {Wood-Vasey}, {Allen}, {Garnavich},
  {Jha}, {Joyce}  \& {Matheson}}{{Weyant} et~al.}{2014}]{dist23}
{Weyant} A.,  {Wood-Vasey} W.~M.,  {Allen} L.,  {Garnavich} P.~M.,  {Jha}
  S.~W.,  {Joyce} R.,   {Matheson} T.,  2014, \mn@doi [\apj]
  {10.1088/0004-637X/784/2/105}, \href
  {http://adsabs.harvard.edu/abs/2014ApJ...784..105W} {784, 105}

\bibitem[\protect\citeauthoryear{{Wood-Vasey} et~al.,}{{Wood-Vasey}
  et~al.}{2008}]{dist17}
{Wood-Vasey} W.~M.,  et~al., 2008, \mn@doi [\apj] {10.1086/592374}, \href
  {http://adsabs.harvard.edu/abs/2008ApJ...689..377W} {689, 377}

\bibitem[\protect\citeauthoryear{{Yoon}}{{Yoon}}{2015}]{yoon}
{Yoon} S.-C.,  2015, \mn@doi [\pasa] {10.1017/pasa.2015.16}, \href
  {http://adsabs.harvard.edu/abs/2015PASA...32...15Y} {32, 15}

\bibitem[\protect\citeauthoryear{{Zezas}, {Ward}  \& {Murray}}{{Zezas}
  et~al.}{2003}]{2005U_zezas}
{Zezas} A.,  {Ward} M.~J.,   {Murray} S.~S.,  2003, \mn@doi [\apjl]
  {10.1086/378144}, \href {http://adsabs.harvard.edu/abs/2003ApJ...594L..31Z}
  {594, L31}

\bibitem[\protect\citeauthoryear{{Zezas}, {Fabbiano}, {Baldi}, {Schweizer},
  {King}, {Ponman}  \& {Rots}}{{Zezas} et~al.}{2006}]{2004gt_catalog_source}
{Zezas} A.,  {Fabbiano} G.,  {Baldi} A.,  {Schweizer} F.,  {King} A.~R.,
  {Ponman} T.~J.,   {Rots} A.~H.,  2006, \mn@doi [\apjs] {10.1086/501526},
  \href {http://adsabs.harvard.edu/abs/2006ApJS..166..211Z} {166, 211}

\bibitem[\protect\citeauthoryear{{Zhekov}, {Gagn{\'e}}  \& {Skinner}}{{Zhekov}
  et~al.}{2011}]{zhekov}
{Zhekov} S.~A.,  {Gagn{\'e}} M.,   {Skinner} S.~L.,  2011, \mn@doi [\apjl]
  {10.1088/2041-8205/727/1/L17}, \href
  {http://adsabs.harvard.edu/abs/2011ApJ...727L..17Z} {727, L17}

\makeatother
\end{thebibliography}

\bsp	
\label{lastpage}
\end{document}